\documentclass[12pt]{article}
\newlength{\abstractwidth}
\renewcommand{\thanks}[1]{\footnote{#1}} 

\newcommand{\be}{\begin{equation}}
\newcommand{\bea}{\begin{eqnarray}}
\newcommand{\eea}{\end{eqnarray}}
\newcommand{\ee}{\end{equation}}

\newcommand{\<}{\langle}
\renewcommand{\>}{\rangle}
\def\ba{\begin{eqnarray}}
\def\ea{\end{eqnarray}}

\def\Im{{\rm Im}}

\def\det{{\rm det}}
\def\half{ {1\over 2}}

\def\14{{1 \over 4}}
\def\p{\partial}
\def\tet{\vartheta}
\def\no{\nonumber}
\def\e{\epsilon}
\def\no{\nonumber}
\def\tet{\vartheta}
\def\ep{\varepsilon}

\begin{document}
\baselineskip=16pt

\begin{flushright}
UCLA/04/TEP/46 \\
Columbia/Math/04 \\
2004 November 17
\end{flushright}

\bigskip

\begin{center}
{\Large \bf  ASYZYGIES, MODULAR FORMS, AND
\\
\medskip
THE SUPERSTRING MEASURE I
\footnote{Research supported in
part by National Science Foundation grants PHY-01-40151 and DMS-02-45371.}}

\bigskip\bigskip

{\large Eric D'Hoker$^a$ and D.H. Phong$^b$} \\ 

\bigskip

$^a$ \sl Department of Physics and Astronomy\\
\sl University of California, Los Angeles, CA 90095 \\
$^b$ \sl Department of Mathematics \\ 
\sl Columbia University, New York, NY 10027

\end{center}

\bigskip\bigskip

\begin{abstract}

The goal of this paper and of a subsequent continuation is
to find some viable ansatze for the three-loop
superstring chiral measure.
For this,
two alternative formulas are derived for the two-loop superstring chiral measure. Unlike the original formula, both alternates  admit 
modular covariant generalizations to higher genus. One of these two generalizations is analyzed in detail in the present paper, with the 
analysis of the other left to the next paper of the series.

\end{abstract}
 
\newpage


\newpage

\baselineskip=15pt
\setcounter{equation}{0}
\setcounter{footnote}{0}
\newtheorem{theorem}{Theorem}
\newtheorem{definition}{Definition}

\vfill\eject

\section{Introduction}
\setcounter{equation}{0}

String perturbation theory is closely related to the mathematical
theories of Riemann surfaces and modular forms.
In string theory, scattering amplitudes are given perturbatively
at loop order $h$ by the moments of a measure on the moduli space of Riemann surfaces of genus $h$. For example, the vacuum-to-vacuum amplitude, or cosmological constant, is given by
an expression of the form
\be
{\cal A}
=
\int_{{\cal M}_h}(\det\, \Im\Omega)^{-5}
\sum_{\delta,\bar\delta}c_{\delta,\bar\delta}
\
d\mu[\delta](\Omega)\wedge \overline{d\mu[\bar\delta](\Omega)}
\ee
where $\delta,\bar\delta$ denote independent spin structures, 
$c_{\delta,\bar\delta}$ are suitable constant phases, and $d\mu[\delta](\Omega)$ is a holomorphic form of maximal rank $(3h-3,0)$ on the moduli space 
of Riemann surfaces of genus $h$.
Here we have represented the Riemann surface by its period matrix $\Omega$, after a choice of canonical homology basis. Since the
integrand in the above formula should be independent of the homology basis, it follows that $d\mu[\delta](\Omega)$ must transform covariantly under 
$Sp(2h,{\bf Z})$ modular transformations.

\medskip

This correspondence between the superstring measure and modular forms is particularly simple at one-loop, where $d\mu[\delta]
\sim\tet[\delta](0,\Omega)^4/\eta(\Omega)^{12}$,
so that $d\mu[\delta](\Omega)$ is essentially a $\tet$-constant. 
No such simple relation is known in general.
It has long been hoped that the extensive mathematical
literature on modular forms, combined with physical constraints
on the superstring measure such as factorization, can help pinpoint the measure. But this hope has remained unfulfilled,
perhaps because of the lack of a suitable Ansatz.

\medskip
In \cite{I, II, III, IV} (see also the review \cite{dp02}), an explicit 
formula for the two-loop superstring measure was derived from 
first principles. It is given by
\be
\label{stringmeasure}
d\mu[\delta]={\tet[\delta]^4(0,\Omega)\,\Xi_6[\delta](\Omega)
\over
\Psi_{10}(\Omega)}\prod_{1\leq I\leq J\leq 2}d\Omega_{IJ}
\ee
Here $\Psi _{10} (\Omega)=\prod_{\delta\ even}\tet[\delta](0,\Omega)^2$ 
is the well-known modular form introduced by Igusa \cite{igusa1, igusa2}, 
and the key expression $\Xi_6[\delta](\Omega)$ is
defined by 
\bea
\label{genus2}
\Xi _6 [\delta ](\Omega) \equiv
\sum _{1 \leq i < j \leq 3} \< \nu_i | \nu_j\>
\prod _{k=4,5,6} \tet [\nu_i + \nu _j + \nu_k]^4(0,\Omega),
\eea 
where we have expressed the even spin structure $\delta$
as the sum of three odd spin structures,
say $\delta=\nu_1+\nu_2+\nu_3$, and $\nu_4,\nu_5,\nu_6$
are the remaining odd spin structures.
This prescription clearly depends heavily on the fact
that the surface has genus $2$.

\medskip

The main goal of the present paper is to 
provide two alternative formulas of $\Xi_6[\delta](\Omega)$ 
which are no longer tied specifically to genus $2$. 
One formula still retains the form
(\ref{genus2}) (so that $\Xi_6[\delta]$ is still expressed
as a linear combination of the 4th-powers of {\it three} $\tet$-constants), 
but the contributing spin structures
$\nu_i+\nu_j+\nu_4$,
$\nu_i+\nu_j+\nu_5$,
$\nu_i+\nu_j+\nu_6$ are now characterized as triples
$\{\delta_m,\delta_n,\delta_p\}$ which together with $\delta$
form a {\it totally asyzygous} quartet (Theorem 1). The other formula
is a more radical departure from (\ref{genus2}), since $\Xi_6[\delta]$ is now expressed as a linear combination
of the squares of {\it six} $\tet$-constants.
The corresponding sextets of spin structures are characterized
by the requirements that they don't contain $\delta$, and can be
divided into three pairs of spin structures, the union of any two
is a totally asyzygous quartet (Theorem 2).

\medskip

The new formula in terms of $4$-th powers of $\tet$-constants
admits immediately a generalization to all genera $h\geq 2$.
The generalization to genus $3$ of the formula in terms of squares of 
$\tet$-constants is already more delicate. We shall describe it 
only in the next paper of this series, and devote the rest of this 
paper to the study of the $4$-th powers formula.   
We show that the generalization of the 4-th powers formula
satisfies the same modular transformation laws as in genus $2$ (Theorem 3)
and the same relation with the basic modular forms
$\Psi_4(\Omega)^2-2^h\Psi_8(\Omega)$, upon summation over
spin structures (Theorem 4). The expression $\Psi_4(\Omega)^2-2^h\Psi_8(\Omega)$ is well-known to vanish in genus $1$ by the Jacobi identity, and in genus $2$ as a consequence of Igusa's theorem on the ring of modular forms.
Our numerical simulations indicate very strongly that it also
vanishes in genus $3$, and a natural conjecture is that
it actually vanishes for all genera. We also determine explicitly
the degeneration limits of the $4$-th powers formula in genus $3$.

\medskip

As we had stressed earlier, our main motivation for
this study is to develop a suitable Ansatz for the superstring measure in higher genus. In the next paper of this series,
we shall show, based on the degeneration properties of the
two formulas, that it is the formula in terms of squares of
$\tet$-constants of sextets of spin structures which leads to
viable candidates for the three-loop superstring measure.
Nevertheless, both formulas exhibit remarkable properties,
and it can be hoped that they will both be of interest in the
theories of Riemann surfaces and modular forms.

\newpage

\section{Alternative formulas for $\Xi_6[\delta](\Omega)$}
\setcounter{equation}{0}

On a Riemann surface $\Sigma$ of genus $h$, there are $2^{2h}$
spin structures, i.e., square roots of the canonical bundle of
$\Sigma$. Fixing a canonical homology basis, a spin structure 
$\delta$ can be identified with a half-characteristic
$\delta=(\delta'|\delta'')$, $\delta',\delta''\in \{0,\half\}^h$.
The $2^{2h}$ spin structures can be divided into 
$2^{h-1}(2^h+1)$ even and $2^{h-1}(2^h-1)$ odd ones, depending
on the parity $p(\delta)={\rm exp}(4\pi i\delta'\cdot\delta'')$. 
The relative signature $\<\delta_1|\delta_2\>$ of a pair of spin structures $\delta_1,\delta_2$ is defined by
\be
\<\delta_1|\delta_2\>
={\rm exp}\,\{4\pi i(\delta_1' \cdot\delta_2''-\delta_1''\cdot\delta_2 ')\}
\ee
The relative signature of two spin structures
$\delta_1,\delta_2$ is {\it not} invariant under modular
transformations (for their precise transformation laws, see Appendix \S A). 
However, for a triplet $\delta,\epsilon,\eta$
of spin structures, we can define the sign
$e(\delta, \epsilon, \eta)$ by
\bea
e(\delta, \epsilon, \eta) 
\equiv 
\<\delta |\epsilon\> \cdot 
\<\epsilon |\eta\> \cdot 
\<\eta|\delta \>
\eea
which does not depend on the ordering of the triplet, and is invariant under 
modular transformations. This can be seen from the  
relation between the parities
of a triplet of spin structures and their relative signature,
which is a direct consequence of their definitions
\be
\label{pe}
p(\delta+\epsilon+\eta)
=p(\delta)p(\epsilon)p(\eta)\,e(\delta,\epsilon,\eta).
\ee

\medskip

In genus $h=2$, there are $10$ even spin structures and $6$ odd ones. 
It is convenient to reserve the notation $\delta$ for even spin structures, 
and denote odd spin structures by $\nu$.
Each even spin structure $\delta$ can be written
as $\delta=\nu_1+\nu_2+\nu_3$, where the $\nu_i$'s are pairwise distinct 
odd spin structures. If $\nu_4,\nu_5,\nu_6$ is the complementary set of 
odd spin structures, then we also have $\delta=\nu_4+\nu_5+\nu_6$, 
and the mapping $\{\nu_1,\nu_2,\nu_3\}\to\delta$ is exactly 2 to 1. 
For each even spin structure $\delta$ in genus $2$,
say $\delta=\nu_1+\nu_2+\nu_3$,
the form $\Xi_6[\delta]$ was defined in \cite{I} by the formula (\ref{genus2}).  
In this section, we provide two more 
formulas for $\Xi_6[\delta](\Omega)$. The first one gives a different 
description of the even spin structures whose $\tet$-constants enter 
the right hand side of (\ref{genus2}). The second one replaces the 
product of 3 fourth powers of $\tet$-constants by the product of 6 
second powers of different $\tet$-constants. Both formulas
do not depend on the decomposition of even spin structures into odd spin structures in genus $2$ and hence allow
extensions to arbitrary genus. 

\medskip

Let $(\delta_1,\cdots,\delta_N)$ be an unordered $N$-tuplet
of $N$ distinct spin structures. We say that the $N$-tuplet is a totally asyzygous $N$-tuplet if
$e(\delta_{i_1},\delta_{i_2},\delta_{i_3})=-1$ for any
subset of three distinct spin structures 
$\delta_{i_1},\delta_{i_2},\delta_{i_3}$ in the $N$-tuplet.

\bigskip

\begin{theorem}
The expression $\Xi_6[\delta](\Omega)$ in (\ref{genus2})
can be rewritten as
\bea
\label{def2}
\Xi _6 [\delta ] (\Omega ) =
- \half \sum _{\delta, \epsilon, \eta, \kappa\ tot.asyz.}
\< \delta | \epsilon\>  \< \delta | \eta\> \< \delta | \kappa\> \cdot \tet [\epsilon] ^4(0,\Omega) \tet [\eta]^4 (0,\Omega)
\tet [\kappa]^4 (0,\Omega)
\eea
Here the summation is over all triples $\epsilon,\eta,\kappa$
which together with $\delta$ form a totally asyzygous quartet. Each triplet $(\epsilon, \eta,\kappa)$ is counted only once, irrespective of its ordering.
\end{theorem}

\noindent
\bigskip
{\it Proof.} We have to show that, given an even spin structure $\delta=\nu_1+\nu_2+\nu_3$, the quartets
\be
\label{quartet}
\{\nu_i+\nu_j+\nu_k\}_{\{3\leq k\leq 6\}},
\ee
are asygygous for any choice of $ij$ among $12$, $13$, $23$,
and that conversely, any totally asyzygous quartet $\{\delta,\epsilon,\eta,\kappa\}$ containing $\delta$
must be of this form (or
of the mirror form $\{\nu_i+\nu_j+\nu_k\}_{\{1\leq k\leq 4\}}$
if $\delta$ is represented as $\delta=\nu_4+\nu_5+\nu_6$).

\medskip
We begin by showing that (\ref{quartet}) is asyzygous.
Recall that in genus $2$, it follows immediately from the description of even and odd spin structures that $\nu_i+\nu_j+\nu_k$ is even if and only if the odd spin structures $\nu_i,\nu_j,\nu_k$ are pairwise distinct.
Thus the spin structures in the quartet in (\ref{quartet}) are all even. 
Furthermore, the sum of any distinct three of
them is $3(\nu_i+\nu_j)+\nu_{k_1}+\nu_{k_2}+\nu_{k_3}
=\nu_i+\nu_j+\nu_{k_1}+\nu_{k_2}+\nu_{k_3}$, and hence equal
to the odd spin structure $\nu_{k_4}$, where $k_4$ is the remaining index in $\{3,4,5,6\}$. It follows now from the
relation (\ref{pe}) that the triplet $\{\nu_i+\nu_j+\nu_k\}_{\{k=k_1,k_2,k_3\}}$ is asyzygous, and hence the quartets (\ref{quartet}) are totally asyzygous.

\medskip

We show now the converse. Since all the spin structures in
$\{\delta,\epsilon,\eta,\kappa\}$ are assumed to be even,
the totally asyzygous condition is equivalent, in view of
the relation (\ref{pe}) to the condition that the sum of any
three distinct spin structures in the quartet be odd. We can easily
work this out for $\delta=\nu_1+\nu_2+\nu_3$. Recall that each even spin structure admits two representations, as the sum of a set of three distinct odd spin structures $\nu'+\nu''+\nu'''$, or as the complementary sum. Since the spin structures $\epsilon,\eta,\kappa$
are different from $\delta$, we may assume that, for each of them, at least one of the three spin structures $\nu',\nu'',\nu'''$ is among $\nu_1,\nu_2,\nu_3$. By
taking the complementary representation if necessary, we may assume that at least two of them are in this set, and in fact
exactly two, since they are distinct from $\delta$.
Thus, after some suitable renaming, the spin structures
$\epsilon,\eta,\kappa$ must fall into one of the following
three possibilities
\be
(1)\ \cases{\epsilon=\nu_1+\nu_2+\nu_j\cr
\eta=\nu_1+\nu_3+\nu_k\cr
\kappa=\nu_2+\nu_3+\nu_l\cr}
\ \
(2)\ \cases{\epsilon=\nu_1+\nu_2+\nu_j\cr
\eta=\nu_1+\nu_2+\nu_k\cr
\kappa=\nu_1+\nu_3+\nu_l\cr}
\ \
(3)\ \cases{\epsilon=\nu_1+\nu_2+\nu_j\cr
\eta=\nu_1+\nu_2+\nu_k\cr
\kappa=\nu_1+\nu_2+\nu_l\cr}
\ee
with $\nu_j,\nu_k,\nu_l\in\{4,5,6\}$.

\medskip

The case (3) gives a quartet of type (\ref{quartet}), as claimed.
To deal with the cases (1) and (2), we again exploit the fact that the 
sum of the spin structures in an asygygous triplet of even spin 
structures must be odd. Applied to a triplet of 
the form $\{\delta=\nu_1+\nu_2+\nu_3,\nu_1+\nu_2+\nu_j,
\nu_1+\nu_3+\nu_k\}$, it implies that $\nu_3+\nu_j+\nu_k$
must be odd. But this implies that $\nu_j=\nu_k$ (since they
are both different from $\nu_3$). Turning more specifically to
the case (1), we deduce that the spin structures $\nu_j,\nu_k,\nu_l$ 
must all be the same in this case. 
Going to the complementary representation for $\{\delta,
\epsilon,\eta,\kappa\}$, we obtain a quartet of the form
(\ref{quartet}), with $ij$ given by one of the pairs $45$, $46$,
or $56$. Finally, for case (2), we note that the previous
observation implies that $\nu_k=\nu_l$. But then we have
$\epsilon+\eta+\kappa=\nu_1+\nu_3+\nu_j$, which is even, 
contradicting the fact that the triplet $\epsilon,\eta,\kappa$
is asyzygous.

\medskip

It remains to re-express the sign factor $\<\nu_i,\nu_j\>$
in (\ref{genus2}) in the form of indicated in (\ref{def2}).
For this, consider an asyzygous quartet  $\{\delta,\epsilon,\eta,\kappa\}$ 
written under the
form corresponding to case (3) above. Then
$\<\delta|\epsilon\>\<\delta|\eta\>\<\delta|\kappa\>
=\<\delta|\epsilon+\eta+\kappa\>=\<\delta|\nu_3\>
=\<\nu_1|\nu_3\>\<\nu_2|\nu_3\>$.
Since the triplet $\{\nu_1,\nu_2,\nu_3\}$ is asyzygous,
this is equal to $-\<\nu_1|\nu_2\>$. Q.E.D.

\bigskip

\begin{theorem}
The form $\Xi_6[\delta](0,\Omega)$ 
can also be expressed as
\be
\label{six}
\Xi_6[\delta](\Omega)
=
{1\over 2}\sum _{\{ \delta _i\}}
\epsilon(\delta;\{\delta_i\})
\prod_{j=1}^6\tet[\delta_{i_j}](0,\Omega)^2,
\ee
where the sum runs over all sextets 
$\{\delta_{i_1},\cdots,\delta_{i_6}\}$ of even spin structures
which are {\it $\delta$-admissible} in the following sense:
{\rm (1)} $\delta\notin\{\delta_{i_j}\}$; and {\rm (2)}
the sextet can be partitioned into 3 disjoint pairs $\{\delta_{i_1},\delta_{i_2}\}$, $\{\delta_{i_3},\delta_{i_4}\}$, $\{\delta_{i_5},\delta_{i_6}\}$, the 
union of any two pairs forming a totally asyzygous quartet.

\medskip

The modular group $Sp(4,{\bf Z}_2)$ acts transitively
on the space of $7$-tuplets $\{\delta;\delta_{i_1},\cdots,\delta_{i_6}\}$
satisfying the condition that
$\{\delta_{i_1},\cdots,\delta_{i_6}\}$ id $\delta$-admissible.
The expressions $\epsilon(\delta;\delta_{i_1},\cdots,\delta_{i_6})\in\{\pm 1\}$
are phases satisfying the following modular transformation law
\be
\epsilon(M\delta;\{M\delta_{i_j}\}) \epsilon(\delta,M)^4
=
\epsilon(\delta;\{\delta_{i_j}\})
\prod_{k=1}^6
\epsilon(\delta_{i_k},M)^2
\ee
where $\epsilon(\delta,M)$ is the $8$-th root of unity entering
transformation laws for $\tet$-constants. The phases
$\epsilon(\delta;\delta_{i_1},\cdots,\delta_{i_6})$
are completely determined by one another by this transformation law. The global relative sign in (\ref{six}) is determined by the
choice $\e(\delta_7;\{\delta_3,\delta_4,\delta_5,
\delta_6,\delta_9,\delta_0\})=-1$,
in the indexing of even spin structures defined in Appendix
\S B. Their explicit form will be given below,
in (\ref{sign1}) and (\ref{sign2}).
\end{theorem}

\bigskip

\noindent
{\it Proof.} The main tool in the proof is the Riemann relations
for $\tet$-functions. Riemann relations are usually applied to 4th-powers of $\tet$-constants, but they apply equally well to 2nd powers, as we show now. Recall that the Riemann relations
are given by \cite{IV}, \S 2.4 
\bea
\sum _\lambda \< \kappa |\lambda \> 
\tet [\lambda ] (\zeta _1) \tet [\lambda ] (\zeta _2)
 \tet [\lambda ] (\zeta _3) \tet [\lambda ] (\zeta _4)
 = 4 \prod _{i=1}^4 \tet [\kappa ] (\zeta _1 ')
 \eea
where $\zeta '$ is expressed as linear combinations of the $\zeta$.
The case of interest here is for $\zeta _1 = \zeta _2 =\ep$ and 
$\zeta _3 = \zeta _4 =0$, where $\ep\not=0$ is a ``twist" half-characteristic.
Thus, the relations reduce to 
$\zeta _1 ' = \zeta _2 ' = \ep$ and $\zeta _3 ' = \zeta _4 '=0$.
Restricting attention to $\kappa = \nu$ odd, we are left with
\bea
0= 
\sum _\delta \< \delta | \nu \> \tet [\delta ](\ep)^2 \tet [\delta ](0)^2 
=
\sum _\delta e^{- 4 \pi i \ep ' \delta ''} 
\< \delta | \nu \> \tet [\delta + \ep ](0)^2 \tet [\delta ](0)^2 
\eea
This sum receives non-trivial contributions only from terms with both 
$\delta$ and $\delta + \ep$ even.  Any of the 15  twists may be expressed 
(see for example \cite{Aoki:2003sy})
as the sum of two odd spin structures, $\ep = \nu _a + \nu _b$
for $a\not= b$, all such spin structures may be parametrized by
\bea
\delta = \nu _a + \nu _j + \nu _k, \qquad
\delta + \ep = \nu _b + \nu _j + \nu _k 
\hskip 1in  j,k \not \in \{ a,b \}
\eea
Of the 6 possible odd spin structures, 4 yield trivially 0.
The two (equivalent) non-trivial cases are for $\nu = \nu _a$ and $\nu = \nu _b$.
Choosing the first, the Riemann identity may now be recast in the following form,
\bea
0= \sum _{{ j \not= k \atop j,k \not= a,b}}
\varphi (ab|jk) \tet [\nu _a + \nu _j + \nu _k](0)^2
\tet [\nu _b + \nu _j + \nu _k](0)^2
\eea
where 
\bea
\varphi (ab | jk) = \varphi (ba |jk) = \exp 4 \pi i \{ \nu _a '' (\nu _j ' + \nu _k ')
+ \nu _b ' (\nu _j '' + \nu _k '') \}
\eea
Here, the symmetry in $a \leftrightarrow b$ is readily shown using the fact that $j \not= k$ and that $a,b \not \in \{j,k\}$.
This form of Riemann identities is closely related to Riemann
identities on Prym varieties, which appeared in the study of the
cosmological constant in certain orbifold models with broken
supersymmetry \cite{Aoki:2003sy, KKS}.

\medskip

We can return now to the proof of Theorem 2 proper.
The property (1) of $\delta$-admissible sextets described in Theorem 2 is clearly invariant under modular transformations. Both properties (1) and (2) are invariant under the subgroup of modular transformations which leave the spin structure $\delta$ unchanged. The transitivity of this subgroup can be read off from the tables of sextets and modular transformations in Appendix \S B. The transformation law for the phases $\e(\delta;\{\delta_{i_j}\})$ implies that
both sides of the equation (\ref{six}) transform in the same way under modular transformations. Thus it suffices to prove the theorem for some fixed even spin structure, say $\delta=\nu_1+\nu_2+\nu_3$. This is $\delta_7$ in the notation of Appendix \S B. In this case, the sextets which
satisfy the properties (1) and (2) are given by
(see table in Appendix \S C)
\be
\label{six3}
\{34 5960\},
\quad
\{192638\},
\quad
\{152048\},
\quad
\{125690\},
\quad
\{134589\},
\quad
\{234680\}.
\ee
where we have abbreviated the sextet $\{\delta_{i_1},\cdots,\delta_{i_6}\}$ by $\{i_1\cdots i_6\}$.

\medskip

We rewrite now $\Xi_6 [ \nu _1 + \nu _2 + \nu_3]$
using the preceding version of Riemann identities with squares
of $\tet$-constants.
Using the shorthand $(ijk) = \tet [\nu _i + \nu _j + \nu _k](0)^2$, we have 
the following Riemann relations needed to recast 
$\Xi_6 [ \nu _1 + \nu _2 + \nu_3]$,
\bea
\label{Riemann}
\varphi (45 | 12) ~ (124) (125) 
& = & 
- \varphi (45 | 13) ~(134) (135)
- \varphi (45 | 23) ~(234) (235)
\no \\
\varphi (56 | 12) ~(125) (126) 
& = & 
- \varphi (56 | 13) ~(135) (136)
- \varphi (56 | 23) ~(235) (236)
\no \\
\varphi (64 | 12) ~ (124) (126) 
& = & 
- \varphi (64 | 13) ~(134) (136)
- \varphi (64 | 23) ~(234) (236) \qquad
\eea
Multiplying the three left hand sides together and the three right hand sides
together, and using the identity
\bea
\varphi (45 | ij) \varphi (56 | ij) \varphi (64 | ij) = - \< \nu _i |\nu _j \>
\eea
we readily see that 
\bea
\Xi _6 [\nu _1 + \nu _2 + \nu _3] & = & 
- \varphi (45 | 13) \varphi (56 | 13) \varphi (64 | 23) ~ 
(134)(135)^2(136)(234)(236)
\no \\
&&
- \varphi (45 | 13) \varphi (56 | 23) \varphi (64 | 13) ~ 
(134)^2(135)(136)(235)(236)
\no \\
&&
- \varphi (45 | 23) \varphi (56 | 13) \varphi (64 | 13) ~ 
(134)(135)(136)^2(234)(235)
\no \\
&&
- \varphi (45 | 23) \varphi (56 | 13) \varphi (64 | 23) ~ 
(135)(136)(234)^2 (235)(236)
\no \\
&&
- \varphi (45 | 23) \varphi (56 | 23) \varphi (64 | 13) ~ 
(134)(136)(234)(235)^2(236)
\no \\
&&
- \varphi (45 | 13) \varphi (56 | 23) \varphi (64 | 23) ~ 
(134)(135)(234)(235)(236)^2
\eea
Combining lines 1 and 4, lines 2 and 6, and lines 3 and 5
using the same Riemann identities (\ref{Riemann}), we get
\bea
\label{six2}
\Xi _6 [\nu _1 + \nu _2 + \nu _3] & = & 
- c_A (124)(125)(135)(136)(234)(236)
\no \\
&&
- c_B (125)(126)(134)(136)(234)(235)
\no \\
&&
- c_C (124)(126)(134)(135)(235)(236) \qquad
\eea
The signs are easily simplified, and we get
\bea
c_A & = & 
\varphi (45 | 12) \varphi (56 | 13) \varphi (64 | 23)
=
 \exp 4 \pi i \{ \nu _1 ' \nu _6 '' + \nu _2 ' \nu _5 '' + \nu _3 ' \nu_4 ''
+ \nu _4 ' \nu _1 '' + \nu _5 ' \nu _3 '' + \nu _6 ' \nu _2 '' \}
\no \\
c_B & = & \varphi (45 | 23) \varphi (56 | 12) \varphi (64 | 13)
=
 \exp 4 \pi i \{ \nu _2 ' \nu _6 '' + \nu _3 ' \nu _5 '' + \nu _1 ' \nu_4 ''
+ \nu _4 ' \nu _2 '' + \nu _5 ' \nu _1 '' + \nu _6 ' \nu _3 '' \}
\no \\
c_C & = & 
\varphi (45 | 12) \varphi (56 | 13) \varphi (64 | 23)
=
 \exp 4 \pi i \{ \nu _3 ' \nu _6 '' + \nu _1 ' \nu _5 '' + \nu _2 ' \nu_4 ''
+ \nu _4 ' \nu _3 '' + \nu _5 ' \nu _2 '' + \nu _6 ' \nu _1 '' \} 
\no \\
\eea
To re-express the above result in terms of even spin structures, we need to choose a
basis for the even spin structures in terms of the odd ones.
We choose the same basis as was used in \cite{IV}, but the actual values
of the odd spin structures $\nu _a$ are left arbitrary. Modular 
transformations simply permute the odd spin structures.
The basis may be expressed as follows (up to complete periods),
\bea
\delta _1 = \nu _1 + \nu _4 + \nu _6 ~=~ \nu _2 + \nu _3 + \nu _5 & \hskip .5in &
\delta _6 = \nu _1 + \nu _5 + \nu _6 ~=~ \nu _2 + \nu _3 + \nu _4
\no \\
\delta _2 = \nu _1 + \nu _2 + \nu _6 ~=~ \nu _3 + \nu _4 + \nu _5
& \hskip .5in &
\delta _7 = \nu _1 + \nu _2 + \nu _3 ~=~ \nu _4 + \nu _5 + \nu _6
\no \\
\delta _3 = \nu _1 + \nu _2 + \nu _5 ~=~ \nu _3 + \nu _4 + \nu _6
& \hskip .5in &
\delta _8 = \nu _1 + \nu _3 + \nu _4 ~=~ \nu _2 + \nu _5 + \nu _6
\no \\
\delta _4 = \nu _1 + \nu _4 + \nu _5 ~=~ \nu _2 + \nu _3 + \nu _6
& \hskip .5in &
\delta _9 = \nu _1 + \nu _3 + \nu _6 ~=~ \nu _2 + \nu _4 + \nu _5
\no \\
\delta _5 = \nu_1 + \nu _2 + \nu _4 ~=~ \nu_3 + \nu _5 + \nu_6
& \hskip .5in &
\delta _0 = \nu _1 + \nu _3 + \nu _5 ~=~ \nu _2 + \nu _4 + \nu _6 \qquad
\eea
In total, there are 15 totally asyzygous quartets, 9 of which do not
contain a given spin structure. The 9 asyzygous quartets which
do not contain, say $\delta _7$, may be grouped in terms of 
two groups of three $\delta_7$-admissible sextets, given as follows,
\bea
A =  [34 ~ 59 ~ 60 ]   & \hskip .6in & A' = [18 ~ 39 ~ 45]
\no \\
B =  [19 ~ 26 ~ 38 ]  & \hskip .6in & B' = [28 ~ 36 ~ 40]
\no \\
C =  [15 ~ 20 ~ 48]   & \hskip .6in & C' = [12 ~ 50 ~ 69]
\eea
The structure of the asyzygous quartets allows us to recast
odd spin structures in terms of asyzygous triplets of even ones,
for each group, as follows in Table 1 below.

\begin{table}[htdp]
\begin{center}
\begin{tabular}{|c||c|c|c||c|c|c|} 
\hline
$\nu $ & A & B & C & A' & B' & C' \\
\hline \hline
$\nu_1$ & 
$\delta _3 + \delta _4 + \delta _5$ &
$\delta _1 + \delta _8 + \delta _9$ &
$\delta _4 + \delta _8 + \delta _0$ &
$\delta _1 + \delta _8 + \delta _9$ &
$\delta _2 + \delta _3 + \delta _6$ &
$\delta _1 + \delta _2 + \delta _5$ 
\\ \hline
$\nu_2$ & 
$\delta _3 + \delta _5 + \delta _9$ &
$\delta _1 + \delta _6 + \delta _9$ &
$\delta _1 + \delta _4 + \delta _8$ &
$\delta _1 + \delta _4 + \delta _8$ &
$\delta _2 + \delta _3 + \delta _8$ &
$\delta _1 + \delta _6 + \delta _9$ 
\\ \hline
$\nu_3$ & 
$\delta _3 + \delta _4 + \delta _6$ &
$\delta _1 + \delta _2 + \delta _6$ &
$\delta _1 + \delta _4 + \delta _5$ &
$\delta _1 + \delta _4 + \delta _5$ &
$\delta _2 + \delta _8 + \delta _0$ &
$\delta _1 + \delta _2 + \delta _6$ 
\\ \hline
$\nu_4$ & 
$\delta _3 + \delta _6 + \delta _0$ &
$\delta _1 + \delta _3 + \delta _8$ &
$\delta _2 + \delta _4 + \delta _8$ &
$\delta _1 + \delta _3 + \delta _8$ &
$\delta _2 + \delta _4 + \delta _8$ &
$\delta _1 + \delta _5 + \delta _0$ 
\\ \hline
$\nu_5$ & 
$\delta _3 + \delta _4 + \delta _9$ &
$\delta _1 + \delta _2 + \delta _9$ &
$\delta _2 + \delta _4 + \delta _0$ &
$\delta _1 + \delta _5 + \delta _8$ &
$\delta _2 + \delta _4 + \delta _0$ &
$\delta _1 + \delta _2 + \delta _9$ 
\\ \hline
$\nu_6$ & 
$\delta _3 + \delta _4 + \delta _0$ &
$\delta _1 + \delta _3 + \delta _9$ &
$\delta _4 + \delta _5 + \delta _8$ &
$\delta _1 + \delta _3 + \delta _9$ &
$\delta _2 + \delta _6 + \delta _8$ &
$\delta _1 + \delta _2 + \delta _0$ 
\\ \hline
\end{tabular}
\caption{Decomposition of odd into even within a $\delta_7$-admissible sextet}
\end{center}
\end{table}

Re-expressing the signs $c_A, c_B, c_C$,  in terms of even spin structures,
we obtain,
\bea
\label{sign1} 
\Xi _6 [\delta _7 ] 
& = & 
- \psi (\delta _4, \delta _5, \delta _0) ~ \psi (\delta _3, \delta _9, \delta _6) ~
\tet [\delta _3]^2 \tet [\delta _4]^2 \tet [\delta _5]^2 
\tet [\delta _6]^2 \tet [\delta _9]^2 \tet [\delta _0]^2 
\no \\ &&
- \psi (\delta _3, \delta _9, \delta _6) ~ \psi (\delta _1, \delta _2, \delta _8) ~
\tet [\delta _1]^2 \tet [\delta _2]^2 \tet [\delta _3]^2 
\tet [\delta _6]^2 \tet [\delta _8]^2 \tet [\delta _9]^2 
\no \\ &&
- \psi (\delta _1, \delta _2, \delta _8) ~ \psi (\delta _4, \delta _5, \delta _0) ~
\tet [\delta _1]^2 \tet [\delta _2]^2 \tet [\delta _4]^2 
\tet [\delta _5]^2 \tet [\delta _8]^2 \tet [\delta _0]^2 \qquad
\eea
where the following combination enters,
\bea
\label{sign2}
\psi (\delta _i, \delta _j, \delta_k)
\equiv
\exp  4 \pi i \bigg \{
\delta _i ' \delta _j '' + \delta _j ' \delta _k '' + \delta _k ' \delta _i '' \bigg \}
\eea
and our final expression for $\Xi _6 [\delta _7]$. An alternative formula
is obtained by replacing the partitioning $A,B,C$ by $A',B',C'$.
Averaging over the two formulas gives the
desired formula (\ref{six}). Note that it is the full set
of sextets in (\ref{six3}) which is an orbit under the
group of modular transformations fixing $\delta$.
Note also that the above expression in terms of even spin structures 
is uniquely defined by the partitioning. This may be established 
using the symmetries of the function $\psi$.
Computing the effect of a permutation, we find
\bea
\psi (\delta _i, \delta _j, \delta_k) \psi (\delta _j, \delta _i, \delta_k)
=
\<\delta _i |\delta _j\> \< \delta _j | \delta _k \> \< \delta _k | \delta _i \>
=
e(\delta _i, \delta _j, \delta _k)
\eea
Next, by inspection, we see that the triplets $(\delta _i, \delta _j, \delta _k)$ 
on which $\psi$ is evaluated in each partition, namely
$(\delta_4, \delta _5, \delta _0)$, $(\delta _3, \delta _9, \delta _6)$,
and $(\delta _1, \delta _2, \delta _8)$, are all 
syzygous, so that $\psi (\delta _i, \delta _j, \delta _k)$ is totally symmetric in 
its three arguments. Q.E.D.

\bigskip
\noindent
{\bf Remarks}

\medskip
$\bullet$ We note that, unlike in the expression for $\Xi_6[\delta]$ in terms of 4th-powers of $\tet$-constants, 
the phases $\e(\delta;\{\delta_i\})$ in the expression (\ref{six}) cannot be expressed in terms of simple modular invariant quantities such as the relative signature of two
spin structures. This will account for some important subtleties
when we discuss the generalization of expressions such as
(\ref{six}) to higher genus in \cite{asyzygyII}.

\medskip
$\bullet$ Another proof of Theorem 2 can be obtained, using the
hyperelliptic representation $s^2=\prod_{i=1}^6(x-p_i)$ of genus 2 Riemann surfaces. If we view the odd spin structures $\nu_i$
as corresponding to the branch points $p_i$, then the spin structure $\delta=\delta_7$ corresponds to the partition of the 6 branch points into two sets $A=\{p_1,p_2,p_3\}$ and $B=\{p_4,p_5,p_6\}$. We can then establish the following formula
\bea
{\Xi_6[\delta] \over \tet[\delta]^2 \Psi_{10}^{1/2}}
=
{ \sum _\sigma (p_1-p_{\sigma (4)} )(p_2-p_{\sigma (5)})(p_3-p_{\sigma (6)})
\over
\sqrt{\prod_{i\not=j\in A}(p_i-p_j)\prod_{i\not=j\in B}(p_i-p_j)}}
\eea
where $\sigma$ runs over the 3 cyclic permutations of $\{4,5,6\}$.
Expressing the right hand side in terms of $\tet$-constants gives again 
(\ref{six2}), up to an overall sign, which cannot be determined from the 
hyperelliptic representation.

\medskip

$\bullet$ We stress that the full orbit under the subgroup of modular transformations leaving $\delta_7$ unchanged requires all
6 sextets listed in (\ref{six3}). However, the fact that this orbit can be 
broken into two groups, each already giving essentially $\Xi_6[\delta]$, 
plays an important role in the
construction in \cite{asyzygyII} of modular forms using pairs of sextets.

\newpage

\section{A modular covariant form $\Xi_6^\#[\delta](\Omega)$
in any genus}
\setcounter{equation}{0}

In the previous section, we have seen that the
modular covariant form $\Xi_6[\delta](\Omega)$ 
admits an expression in terms of totally asyzygous quartets
of even spin structures. This notion makes sense for all 
genera, and thus we can define an extension to all genera of
$\Xi_6[\delta](\Omega)$, denoted by
$\Xi_6^\#[\delta](\Omega)$, by the right hand side of (\ref{def2}). 
The main goal of the present section is to
show that $\Xi_6^\#[\delta](\Omega)$ satisfies modular transformation laws 
and identities similar to those of $\Xi_6[\delta](\Omega)$ in genus $2$. But before we do so, we wish to point out that the
notion of totally asyzygous odd spin structures could have been
used as well.

\bigskip
Let $\{\delta_1,\delta_2,\delta_3,\delta_4\}$ be any quartet of spin
structures, and define another quartet of spin structures $\{\nu_1,\nu_2,\nu_3,\nu_4\}$ by
\be
\nu_i=\sum_{j\not=i}\delta_j,
\ \ \ 1\leq i\leq 4.
\ee
The two quartets determine each other uniquely, since the above equations imply that $\delta_j=\sum_{i\not=j}\nu_i$.
We claim now that $\{\delta_1,\delta_2,\delta_3,\delta_4\}$
is an asyzygous quartet of even spin structures if and only
if $\{\nu_1,\nu_2,\nu_3,\nu_4\}$ is an asyzygous quartet of odd
spin structures. Indeed, assume that $\{\delta_1,\delta_2,\delta_3,\delta_4\}$
is an asyzygous quartet of even spin structures, and consider say $p(\nu_1)$. Then $p(\nu_1)=p(\delta_2)p(\delta_3)p(\delta_4)
e(\delta_2,\delta_3,\delta_4)
=-1$ which shows that $\nu_1$, and hence any $\nu_i$, is  odd. Reversing the roles of $\delta_i$ and $\nu_i$ in the same formula shows that $p(\delta_1)=p(\nu_2)p(\nu_3)p(\nu_4)e(\nu_2,\nu_3,\nu_4)$,
which implies that $e(\nu_2,\nu_3,\nu_4)=-1$, since the $\delta_i$'s are even, and the $\nu_i$'s are now known to be odd.
Thus the quartet $\{\nu_1,\nu_2,\nu_3,\nu_4\}$ is asyzygous.
The same argument can clearly be reversed to show that
$\{\delta_1,\delta_2,\delta_3,\delta_4\}$ is an asyzygous  quartet of even spin structures if $\{\nu_1,\nu_2,\nu_3,\nu_4\}$
is an asyzygous quartet of odd spin structures, completing the
proof of our claim.

\medskip

In view of the above exact correspondence between asyzygous 
quartets of even and odd spin structures, we obtain immediately 
the following equivalent characterization of $\Xi_6^\#[\delta](\Omega)$ 
valid in any genus
\bea
\Xi _6^\#[\delta] 
=  
{1 \over 6} \sum _{[\nu_1, \nu_2, \nu_3, \nu_4]} \! \! \!
\theta_{[\nu_1, \nu_2, \nu_3, \nu_4]} [\delta] ~
\< \delta |\delta _{\sigma(1)} \> 
\< \delta |\delta _{\sigma (2)} \> 
\< \delta |\delta _{\sigma (3)} \> 
\tet [\delta_{\sigma (1)}]^4 \tet [\delta_{\sigma (2)}]^4 \tet [\delta_{\sigma (3)}]^4
\eea
Here the summation is over 
all totally asyzygous quartets  $[\nu_1, \nu_2, \nu_3, \nu_4]$
of odd spin structures. The indicator function is defined by
\bea
\theta _{[\nu_1, \nu_2, \nu_3, \nu_4]} [\delta]
\equiv \left \{
\matrix{ 
1 & {\rm iff} & \delta = \nu_{\sigma(1)} +  \nu_{\sigma(2)} + \nu_{\sigma(3)} \cr
0 & {\rm iff} & \delta \not= \nu_{\sigma(1)} +  \nu_{\sigma(2)} + \nu_{\sigma(3)} \cr}
\right .
\eea
where $\sigma$ is any permutation of $1,2,3,4$.

\medskip

We turn now to the modular properties of $\Xi_6^\#[\delta](\Omega)$:

\bigskip

\begin{theorem}
Under modular transformations $M$, the expression
$\Xi_6^\#[\delta](\Omega)$ transforms as
\bea
\label{modular}
\Xi_6^\# [\tilde \delta ] (\tilde \Omega) = \epsilon (\delta , M)^4
\det (C\Omega +D)^6 \Xi _6^\# [\delta ](\Omega)
\eea
where $\tilde\epsilon=M\epsilon$
and $\epsilon(\delta,M)$ is precisely the same $8$-th root
of unity occurring in the modular transformation for
$\tet[\delta](0,\Omega)$.
\end{theorem}

\bigskip

\noindent
{\it Proof.} We establish (\ref{modular}) for general modular 
transformations by establishing it for the generators of $Sp(2h,{\bf Z})$. 
A set of generators is given by the matrices $M_A$, $S$, and $M_B$ 
listed in Appendix \S A.
Modular transformations map totally asyzygous quartets into totally 
asyzygous quartets.
Now the signature on pairs is invariant under $M_A$ and $S$, 
and the factors $\epsilon(\delta,M_A)$ and $\epsilon(\delta,S)$
are essentially trivial in this case, so we need only consider
the case of $M_B$ generators. We
proceed as follows,
\bea
\Xi _6^\# [\tilde \delta ] (\tilde \Omega ) = - \half 
\sum _{[\tilde \delta, \tilde \epsilon, \tilde \eta, \tilde \kappa]}
\< \tilde \delta | \tilde \epsilon\>  
\< \tilde \delta | \tilde \eta\> 
\< \tilde \delta | \tilde \kappa\>
\cdot
\tet [\tilde \epsilon] ^4(0,\tilde \Omega) 
\tet [\tilde \eta]^4 (0,\tilde \Omega)
\tet [\tilde \kappa]^4 (0,\tilde \Omega)
\eea
Next, we cast each spin structure as the transformation of its pre-image,
and use the $\tet$-constant transformation laws,
\bea
\Xi _6^\# [\tilde \delta ] (\tilde \Omega ) &=& - \half 
\sum _{[ \delta, \epsilon,  \eta, \kappa]}
\< \tilde \delta | \tilde \epsilon\>  
\< \tilde \delta | \tilde \eta\> 
\< \tilde \delta | \tilde \kappa\> \,
\epsilon ^4 (\epsilon, M_B) 
\epsilon ^4 (\eta, M_B)
\epsilon ^4 (\kappa, M_B)
\nonumber \\ && \hskip 1in \times
\tet [ \epsilon] ^4(0, \Omega) 
\tet [ \eta]^4 (0, \Omega)
\tet [ \kappa]^4 (0, \Omega)
\eea
The signature factors worked out to be
\bea
\< \tilde \delta | \tilde \epsilon\>  
\< \tilde \delta | \tilde \eta\> 
\< \tilde \delta | \tilde \kappa\>
=
\< \delta | \epsilon\>  \< \delta | \eta\> \< \delta | \kappa\>
\exp \biggl \{  2 \pi i 
\sum _{i=1} ^h (3 \delta _i' -\epsilon _i ' - \eta _i ' - \kappa _i ') B_i
\biggr \}
\nonumber \\
\epsilon ^4 (\epsilon, M_B) 
\epsilon ^4 (\eta, M_B)
\epsilon ^4 (\kappa, M_B)
= \exp \biggl \{ - 4 \pi i \sum _{i=1} ^h ((\epsilon '_i)^2 + (\eta
'_i)^2 +(\kappa _i ')^2 )B_i \biggr \}
\eea
Assembling both factors, and using the fact that $B_i$ are integers, we obtain
\bea
&&
\< \tilde \delta | \tilde \epsilon\>  
\< \tilde \delta | \tilde \eta\> 
\< \tilde \delta | \tilde \kappa\>
\epsilon ^4 (\epsilon, M_B) 
\epsilon ^4 (\eta, M_B)
\epsilon ^4 (\kappa, M_B)
 \\ && \nonumber \\
&& = \epsilon ^4 (\delta, M_B) 
\< \delta | \epsilon\>  \< \delta | \eta\> \< \delta | \kappa\>
\nonumber \\
&&
\qquad \times 
\exp \biggl \{ -  2\pi i \sum _{i=1} ^h \biggl ( 
\delta _i '  (2 \delta '_i  +1) +
\epsilon '_i (2\epsilon '_i +1) + 
\eta '_i     (2\eta '_i     +1) +
\kappa _i '  (2 \kappa _i ' +1) 
\biggr ) B_i \biggr \}
\nonumber
\eea
Now, in the last exponential factor, the combinations $\delta _i' (2\delta _i '+1)$ are integers whenever $\delta _i'$ are half-integers, and thus the exponential factor is 1.  Q.E.D.

\medskip

It follows from Theorem 3 that
\bea
\tet [\tilde \delta ]^4 (0,\tilde \Omega)\, 
\Xi_6^\# [\tilde \delta ] (\tilde \Omega) =
\det (C\Omega +D)^8 \tet [\delta ]^4 (0,\Omega)\,
 \Xi _6^\# [\delta ](\Omega)
\eea
and hence the expression $\sum_\delta \Xi_6^\#[\delta]\vartheta[\delta]^4$ 
is a modular form of weight 8. More precisely, define the modular form $\Psi_{4k}(\Omega)$ by
\be
\Psi_{4k}(\Omega)
=\sum_\delta \tet[\delta](0,\Omega)^{8k}
\ee
For genus 2, where a complete classification of the polynomial  ring of modular forms exists \cite{igusa1}, these modular forms 
are subject to relations such as $4\Psi _8 = \Psi_4^2$. In higher
genus, we have not found any analogous discussion in the literature. 
Remarkably, we have the following,

\bigskip

\begin{theorem} 
The form $\Xi_6^\#[\delta](\Omega)$ satisfies the following identity 
in any genus $h\geq 2$
\be
\label{identity}
\sum_\delta \Xi_6^\#[\delta]\vartheta[\delta]^4
=
- 2^{2h-5}(2^h\Psi_8-\Psi_4^2)
\ee
\end{theorem}

\bigskip
\noindent
{\it Proof.} 
In fact, we shall prove a more general result for the following more general
quantity, 
\bea
\label{XiA}
\Xi _6 ^A [\delta] 
& \equiv  & 
- {1 \over 2^5 \cdot 3!}
 \sum _{\delta _1, \delta _2, \delta _3}
 \< \delta |\delta _1\> \< \delta |\delta _2\> \< \delta |\delta _3\>
\tet [\delta _1]^4 \tet [\delta _2]^4 \tet [\delta _3]^4
\no \\ && \hskip 1in \times
\left (1 + a_0 e(\delta_1, \delta _2, \delta _3) \right )
\left (1 + a_1 e(\delta, \delta _2, \delta _3) \right )
\no \\ && \hskip 1in \times
\left (1 + a_2 e(\delta, \delta _1, \delta _3) \right )
\left (1 + a_3 e(\delta, \delta _1, \delta _2) \right )
\eea
Here, $A$ is an array of sign assignments $A= [a_0;a_1,a_2,a_3]$, 
with $a_0,a_1,a_2,a_3 = \pm 1$.  Since
$e(\delta_1, \delta _2, \delta _3)e(\delta, \delta _1, \delta _2)
e(\delta, \delta _1, \delta _3)e(\delta, \delta _2, \delta _3) =+1$, 
it follows that $\Xi _6 ^A$ is non-vanishing only if
$a_0\, a_1 \, a_2 \, a_3=+1$, which we  henceforth assume.
Two $A$'s are equivalent if they differ by a permutation of their last 3 entries.
The sign factor $ \< \delta |\delta _1\> \< \delta |\delta _2\> \< \delta |\delta _3\>$
would appear to involve a specific choice. Actually, 
all other choices are equivalent to it. For example, another modular 
covariant choice would be, 
\bea
 \< \delta |\delta _1\> \< \delta_2 |\delta _3\> 
=
  \< \delta |\delta _1\> \< \delta |\delta _2\> \< \delta |\delta _3\> 
  e(\delta, \delta _2, \delta _3)
=
a_1  \< \delta |\delta _1\> \< \delta |\delta _2\> \< \delta |\delta _3\> 
\eea
But, since $a_1$ is independent of $\delta_i$, 
this choice is equivalent to the one in (\ref{XiA}).
As a special case, we have  $\Xi ^\# _6 [\delta ]= \Xi ^{A_-}_6 [\delta]$ 
for $A_- = [-;-,-,-]$.

\medskip

Using the relation $a_0\, a_1 \, a_2 \, a_3=+1$, 
the form of $\Xi _6^A[\delta]$ may be simplified as follows,
\bea
 \Xi _6 ^A [\delta] 
& = & 
- {1 \over 96}
 \sum _{\delta _1, \delta _2, \delta _3}
 \< \delta |\delta _1\> \< \delta |\delta _2\> \< \delta |\delta _3\>
\tet [\delta _1]^4 \tet [\delta _2]^4 \tet [\delta _3]^4
\left (1 + a_0 e(\delta_1, \delta _2, \delta _3) \right )
\no \\ && \hskip .7in \times
\left (1    + a_1 e(\delta, \delta _2, \delta _3)   
        + a_2 e(\delta, \delta _1, \delta _3)
        + a_3 e(\delta, \delta _1, \delta _2) \right )
\eea
We analyze this sum according to its $a_i$-dependence.
The term independent of $a_i$ is proportional to $\tet [\delta ]^{12}$
by the Riemann identities. In the term proportional to $a_1$,
we use $ \< \delta |\delta _1\> \< \delta |\delta _2\> \< \delta |\delta _3\>
e(\delta, \delta _2, \delta _3) =  \< \delta |\delta _1\>  \< \delta_2 |\delta _3\>$.
The sums over $\delta _2$, $\delta _3$ yield $\Psi _4$ by the Riemann
identities, while the sum over $\delta_1$ yields $\tet [\delta ]^4$. 
In the term proportional to $a_0a_1$, we use 
$$
 \< \delta |\delta _1\> \< \delta |\delta _2\> \< \delta |\delta _3\>
e(\delta_1, \delta _2, \delta _3) e(\delta, \delta _2, \delta _3) 
=  \< \delta_1 |\delta \>  \< \delta_1 |\delta _2 \>  \< \delta_1 |\delta _3 \>
$$
The sums over $\delta _2$ and $\delta _3$ yield $\tet [\delta _1]^8$
and the remaining sum over $\delta _1$ is a sum over $\tet [\delta_1]^{12}$.
In summary, we have the following simplified form, valid for general genus $h$,
\bea
\label{XiA1}
\Xi_6 ^A [\delta] 
& = & 
- {1 \over 3} \, 2^{3h-5} (1 + a_0) \tet [\delta ]^{12} 
- {1 \over 3}\,  2^{2h-5} a \left \{ \tet [\delta ]^4 \Psi _4 
+
a_0  \sum _{\delta _1} \< \delta |\delta _1\> \tet [\delta _1]^{12} \right \}
\no \\ && - {a_0 \over 48}
\sum _{\delta _1, \delta _2, \delta _3} 
\bigg (1- e(\delta_1, \delta _2, \delta _3) \bigg )
     \< \delta |\delta _1\> \< \delta |\delta _2\> \< \delta |\delta _3\>
     \tet [\delta _1]^4 \tet [\delta _2]^4 \tet [\delta _3]^4 
\eea
where we use the notation $a=a_1 + a_2 + a_3$.

\medskip

The sum $\sum _\delta \Xi _6 ^A [\delta] \tet [\delta]^4$ may be calculated in 
terms of $\Psi _4$ and $\Psi_8$ using the Riemann relations, in addition
to one further result, which involves the last term in (\ref{XiA1}).  
Concentrating on the  summation over $\delta$, we have to carry out 
the following sum,
\bea
\label{deltasum}
\sum _\delta \< \delta |\delta _1\> \< \delta |\delta _2\> \< \delta |\delta _3\>
\tet [\delta](0)^4 
= 
\sum _\delta \< \delta |\delta _1 + \delta _2 + \delta _3\> \tet [\delta](0)^4
\eea
Using the fact that the triplet $(\delta_1, \delta_2, \delta_3)$ is  
restricted to be asyzygous by the projection prefactor in the last term of (\ref{XiA1}), the spin structure $\delta _1 + \delta _2 + \delta _3$ must be 
odd, and therefore, using the Riemann identities on the $\delta$-sum,
(\ref{deltasum}) cancels for all $h\geq 2$. The final expression is then
\bea
\sum _\delta \Xi _6 ^A [\delta] \tet [\delta]^4
=
- {1 \over 3} \, 2^{3h-5} (1 + a_0)(1 + a) \Psi_8 
- {1 \over 3}\,  2^{2h-5} a  ( \Psi _4 ^2 - 2^h \Psi _8)
\eea
In the special case of $A=A_-$, we have $a_0=-1$ and $a=-3$, and 
we readily recover the statement of Theorem 3. Q.E.D.

\medskip

Note that with the help of $\Xi_6 ^\# [\delta]$, the last sum in (\ref{XiA1}) may
be eliminated, so that
\bea
\Xi ^A _6 [\delta] 
& = & 
- a_0 \Xi_6 ^\#[\delta] 
- {1 \over 3} \cdot 2^{3h-5} (1 + a_0) \tet [\delta ]^{12} 
+  2^{2h-5} (a_0 - {1 \over 3} a) \tet [\delta ]^4 \Psi _4
\no \\ && 
- {1 \over 3} \cdot 2^{2h-5} a_0 ( a +3 ) \sum _{\delta _1} 
\< \delta |\delta _1\> \tet [\delta_1 ]^{12}.
\eea

\newpage

\section{A conjectured identity between $\Psi_8(\Omega)$ and $\Psi_4(\Omega)$}
\setcounter{equation}{0}

The identity in Theorem 3 suggests that the following relation
between $\Psi_8(\Omega)$ and $\Psi_4(\Omega)$ should hold in any genus
\be
\label{conjecture}
2^h\Psi_8(\Omega)-\Psi_4^2(\Omega)=0
\ee
We note that $\Psi _{4k} (\Omega ) \to 2^h$ for all $k$
as $\Im (\Omega _{II}) \to +i \infty$. This number simply corresponds to the $2^h$ spin structures
for which $\delta '_I =0$. This explains why the factor
of $2^h$ in the relation between $\Psi _4$ and $\Psi _8$ is needed. In genus $1$, the relation
(\ref{conjecture}) corresponds to the well-known isospectrality of the root lattices of $E_8\times E_8$ and $SO(32)$,
and follows immediately
from the Jacobi identity $\tet_3^4=\tet_2^4+\tet_4^4$. In genus $2$, it is a consequence of Igusa's classification of genus $2$
modular forms.
To the best of our knowledge, the relation has not been
established for genus $h\geq 3$, although there is an extensive literature on related topics (see e.g. \cite{freitag},
and references therein). We present here some strong
evidence for it in genus $3$. For this, it is useful to define
the following parameters for the period matrix $\Omega$
\bea
\Omega = \left ( \matrix{
\tau _1 & \half \sigma _3 & \half \sigma _2 \cr
\half \sigma _3 & \tau_2  & \half \sigma _1 \cr
\half \sigma _2 & \half \sigma _1 & \tau _3 \cr} \right )
\hskip .8in 
\left \{ \matrix{
x = e^{i \pi \tau_1} \cr
y = e^{i \pi \tau_2} \cr
z = e^{i \pi \tau_3} \cr} \right .
\hskip .8in 
\left \{ \matrix{
u = e^{i \pi \sigma_1} \cr
v = e^{i \pi \sigma_2} \cr
w = e^{i \pi \sigma_3} \cr} \right .
\eea

\subsection{Asymptotic evidence}

The first type of evidence for the relation (\ref{conjecture}) is its vanishing asymptotics under completely separating degenerations. These correspond to $\sigma_i\to 0$. More precisely, we shall show that (\ref{conjecture}) holds for arbitrary $x,y,z$ 
up to and including the orders $\sigma _1 \sigma _2 \sigma _3$ and $\sigma _1 ^2 \sigma _2 ^2, \sigma_2^2\sigma_3^2,
\sigma_1^2\sigma_3^2$.

\medskip
Now the $\tet$-function for the genus $3$ period matrix can be written as
\bea
\tet [\delta ](z,\Omega)
=
\exp \left \{ {1 \over 4 \pi i} \left (
 \sigma _3 \p_1 \p_2 + \sigma _3 \p_1 \p_2 + \sigma _3 \p_1 \p_2 \right ) \right \}
\prod _{i=1,2,3} \tet [\delta_i] (z_i,\tau_i)
\eea
where $\delta^t = (\delta_1 |\delta _2 |\delta _3)^t$, $z=(z_1,z_2,z_3)$ and 
 $\p_i = \p /\p z_i$. 
Expanding the exponential into a series, we see
that even and odd $\delta _i$ behave differently.
We distinguish 2 types of even spin structures $\delta$: those for which
all genus 1 components carry even spin structures $\delta _i$, 
and those for which one component carries an even genus 1 spin structure
but two components carry the odd spin structure. For the latter, $\tet [\delta]$
vanishes to order $\sigma _i \sigma _j$, which implies that the lowest order
contribution to $\Psi _4$ and $\Psi _8$ that is generated by these terms 
is $\sigma _i ^8 \sigma _j^8$. As long as we work to orders less than 8, 
it will thus be necessary to consider only spin structures $\delta$ for which
all genus 1 components $\delta _i$ are even.

\medskip

Taking then all $\delta _i$ to be even and $z_i=0$, and using  the heat equation,
$\p _i ^2 \tet [\delta _i] (z_i,\tau_i) = 4 \pi i \p _{\tau_i} \tet [\delta _i] (z_i,\tau_i)$,
we deduce the following relations,
\bea
\tet [\delta ](0,\Omega) 
& = & 
\sum _{k_i=0 }  ^\infty
{\sigma _1 ^{2k_1} \sigma _2 ^{2k_2} \sigma _3 ^{2k_3} 
\over (2k_1)! (2k_2)! (2k_3)!}
\p _{\tau_1} ^{k_2+k_3} \tet [\delta _1] 
 \p _{\tau_2} ^{k_1+k_3}\tet [\delta _2] 
 \p _{\tau_3} ^{k_1+k_2} \tet [\delta _3] 
 \\ &  & \! +
\sum _{k_i=0 }  ^\infty
{\sigma _1 ^{2k_1+1} \sigma _2 ^{2k_2+1} \sigma _3 ^{2k_3+1} 
\over (2k_1+1)! (2k_2+1)! (2k_3+1)!}
\p _{\tau_1} ^{k_2+k_3+1} \tet [\delta _1] 
 \p _{\tau_2} ^{k_1+k_3+1}\tet [\delta _2] 
 \p _{\tau_3} ^{k_1+k_2+1} \tet [\delta _3] 
\no
 \eea
where we have set 
$\p _{\tau_i} ^k \tet [\delta_i] = \p _{\tau_i} ^k \tet [\delta_i] (0, \tau_i)$. 
To linear order in $\sigma ^1$, the first term occurring is obtained 
from the second term in the sum with $k_1=k_2=k_3=0$. 
Its contribution to $\Psi _4$ and $\Psi_8$ is respectively,
\bea
\Psi _4 
& = & 
\sum _{i,j,k} 
\tet _i ^8 (\tau_1)  \tet _j ^8 (\tau_2) \tet _k ^8 (\tau_3)
\left \{ 1 + 8 \sigma _1 \sigma _2 \sigma _3 
\p_{\tau_1} \ln \tet _i 
\p_{\tau_2} \ln \tet _j 
\p_{\tau_3} \ln \tet _k \right \}
\no \\
\Psi _8 
& = & 
\sum _{i,j,k} 
\tet _i ^{16} (\tau_1)  \tet _j ^{16} (\tau_2) \tet _k ^{16} (\tau_3)
\left \{ 1 + 16 \sigma _1 \sigma _2 \sigma _3 
\p_{\tau_1} \ln \tet _i 
\p_{\tau_2} \ln \tet _j 
\p_{\tau_3} \ln \tet _k \right \}
\eea
Using now the genus 1 relation
$\left (\sum _i \tet _i^8 \right )^2 = 2 \sum _i \tet _i ^{16}$,
we readily see that the equation $\Psi _4^2 = 8 \Psi _8$ is obeyed at 
genus 3 to this order.

\medskip

The next order is of order $\sigma _1 ^2 \sigma _2^2$ (and by symmetry, 
the terms of order $\sigma _1^2 \sigma _3^2$ and $\sigma _2 ^2 \sigma _3^2$
are dealt with in the same way). The contributions to both modular forms are as follows,
\bea
\Psi _4 
& = & 
\sum _{i,j,k} 
\tet _i ^8 (\tau_1)  \tet _j ^8 (\tau_2) \tet _k ^8 (\tau_3)
\left \{ 1 + 4 \sigma _1^2 b_j c_k + 4 \sigma _2^2 a_i c_k +
\sigma _1^2 \sigma _2^2  a_i b_j (2 c_k' + 14 (c_k)^2 
 \right \}
 \\
\Psi _8 
& = & 
\sum _{i,j,k} 
\tet _i ^{16} (\tau_1)  \tet _j ^{16} (\tau_2) \tet _k ^{16} (\tau_3)
\left \{ 1 + 1 + 8 \sigma _1^2 b_j c_k + 8 \sigma _2^2 a_i c_k +
\sigma _1^2 \sigma _2^2  a_i b_j (4 c_k' + 60 (c_k)^2 
 \right \}
\no \eea
with the abbreviations,
$a_i=\p_{\tau_1} \tet _i (\tau_1) /\tet _i (\tau_1)$, 
$b_i =  \p_{\tau_2} \tet _i (\tau_2) /\tet _i (\tau_2)$, 
$c_i =  \p_{\tau_3} \tet _i (\tau_3) /\tet _i (\tau_3)$, 
$c_i '=  \p_{\tau_3}^2 \tet _i (\tau_3) /\tet _i (\tau_3)$. 
Next, we use the following formulas valid in genus $1$,
\be
\sum _i \tet _i (\tau_1)^8 a_i = {1 \over 8} \p_{\tau_1} \Psi _4 (\tau_1)
 \qquad  
\sum _i \tet _i (\tau_1)^{16} a_i = {1 \over 16} \p_{\tau_1} \Psi _8 (\tau_1)
\ee
and similar formulas for $b_i$ and $c_i$.
Putting all together, we have 
\bea
8 \Psi _8 - \Psi _4^2 
& = & {\sigma _1 ^2 \sigma _2 ^2 \over 256}
\Psi _4 (\tau_1)^2 \Psi _4 (\tau_2)^2 \p_{\tau_1} \Psi _4 (\tau_1)
 \p_{\tau_2} \Psi _4 (\tau_2)
\left \{
- 2 \Psi _4 ^2 (\p _{\tau_3} \Psi _4)^2  \right .
 \\
&& \qquad \left . 
+ \sum _k \left [\tet _k (\tau_3)^{16}  (32 c_k ' + 480 (c_k)^2 )
- \tet _k(\tau_3) ^4 \Psi _4 (\tau_3) ( 16 c_k' + 112 (c_k)^2 ) \right ] \right \}
\no
\eea
Finally, using the relation 
\bea
\p_{\tau_3}^2 \left (2 \Psi _8 (\tau_3) - \Psi _4 (\tau_3)^2  \right ) =0
\eea
the above contributions cancel, giving the asymptotics we claimed.

\bigskip

There is also evidence for the relation (\ref{conjecture}) in
a different asymptotic region, namely
in an expansion of $2^h\Psi_8(\Omega)-\Psi_4(\Omega)^2$
in powers of $x,y,z$, but {\sl exact} in $u,v,w$.
A 2hrs calculation using Maple 9 shows that
the relation holds up to and including terms of order 4 in each variable, i.e. up to and including terms 
$x^4, y^4, z^4, x^4y^4, x^4z^4, y^4 z^4$ and $x^4y^4z^4$.

\subsection{Evidence from special cases}

For general $x,y,z,w$ and $u=v=1$ , (or $u=w=1$ or $v=w=1$) the relation follows from the corresponding properties at genus 2, as well as from the relation 
$\Psi _4^2 = 2 \Psi _8$ at genus 1.

\subsection{Numerical evidence}

The genus 3 $\tet$-constants were computed {\sl numerically},
up to and including 50 significant digits at random points in moduli space (this means picking points and verifying that $\Im \Omega >0$).
The Riemann relations are then verified numerically and were found to agree
to approximately 50 digits (recall that there are 36 terms, so a systematic
error would cumulate in the sum and we should expect accuracy to
48 significant digits). Finally, the relations (\ref{conjecture}) was verified numerically. It was found to hold 
within about 48 digits.
This has been carried out for seven points in moduli space; six with purely imaginary $\Omega$ and one with complex $\Omega$.
The values of the points in moduli space were chosen
so that $\Im \Omega >0$. Notice that real $x,y,z,u,v,w$ corresponds to purely imaginary $\Omega$. Numerically, this is the situation when the fewest cancellations occur and the highest precision can be expected.

\bigskip

\begin{table}[htdp]
\begin{center}
\begin{tabular}{|c|c|c|c|c|c||c|c|c|} \hline 
 $x$ & $y$  & $z$ & $u$ & $v$ & $w$ & $\Psi_4$ & $|\Psi _8 -  \Psi _4^2/8|$
                \\ \hline \hline
              $0.3$  
            & $0.4$ 
            & $0.5$        
            & $1.1$ 
            & $1.2$       
            & $1.4$
            & $5.75  \cdot 10^6$ 
            & $ <10^{-35} $

 \\ \hline
              $0.3$  
            & $0.4$ 
            & $0.5$        
            & $1.1$ 
            & $1.2$       
            & $1.3$
            & $4.68 \cdot 10^6$ 
            & $ <10^{-35}$
            
 \\ \hline
             $0.3$  
            & $0.4$ 
            & $0.5$        
            & $1.1$ 
            & $1.2$       
            & $1.2$
            & $3.98 \cdot 10^6$ 
            & $ <10^{-35}$

 \\ \hline
              $0.3$  
            & $0.4$ 
            & $0.5$        
            & $1.1$ 
            & $1.2$       
            & $1.1$
            & $3.56 \cdot 10^6$ 
            & $ <10^{-36}$
            
 \\ \hline
              $0.02$  
            & $0.04$ 
            & $0.05$        
            & $0.11$ 
            & $0.12$       
            & $0.14$
            & $72.1 $ 
            & $ <10^{-47}$
            
 \\ \hline
              $0.0468$  
            & $0.0666$ 
            & $.0734$        
            & $.962$ 
            & $1.11$       
            & $1.25$
            & $7.72$ 
            & $<10^{-46}$
            
 \\ \hline
              $.0468 + .1 i$  
            & $.0666 - .2 i$ 
            & $.0734$        
            & $.962$ 
            & $1.11  - .07 i$       
            & $1.25$
            & $27.6 - 23.6 i$ 
            & $  <10^{-45}$
            
 \\ \hline
\end{tabular}
\end{center}
\caption{Numerical evidence for $\Psi _4 = 8 \Psi _8$ in genus 3}
\label{table:1}
\end{table}

\newpage

\section{Degeneration limits}
\setcounter{equation}{0}

The behavior of a holomorphic modular covariant
form near the divisor of Riemann surfaces with nodes 
is usually of considerable interest. For example,
before the Gliozzi-Scherk-Olive projection,
the superstring measure must exhibit a double pole,
due to the presence of the tachyon (see e.g. 
\cite{pole} and \cite{asyzygyII}
for a detailed discussion). In this section, we shall
analyze the behavior of $\Xi_6^\#$ in the case of genus $3$,
when the underlying surface $\Sigma$ separates into a genus $1$ surface $\Sigma^{(1)}$ and a genus $2$ surface $\Sigma^{(2)}$.

\medskip
Let $(A_I,B_I)$, $1\leq I\leq 3$, be a canonical homology basis for the genus $3$ surface $\Sigma$. Assume that the basis splits up in an
obvious way, with $(A_1, A_2;B_1,B_2)$ associated with the genus two surface while $(A_3, B_3)$ is associated with the genus 1 surface. The
period matrix $\Omega _{IJ}^{(3)}$, $I,J=1,2,3$ may be parametrized as follows,
\bea
\Omega^{(3)} = \left ( \matrix{ \Omega_{11}&\Omega_{12} &  \tau_1 \cr
\Omega_{21} & \Omega_{22}&\tau_2\cr
\tau_1&\tau_2&\Omega_{33}\cr} \right )
\eea
and in the separating limit described above, we have $\tau_1, \tau_2 \to 0$, while $\Omega_{33}$ and  $\{\Omega_{IJ}\}_{1\leq I,J\leq 2}$ tend to the
period matrices $\tau$ and $\Omega$ of $\Sigma^{(1)}$ and $\Sigma^{(2)}$.

\medskip

It is natural to parametrize genus 3 spin structures $\Delta$ following the decomposition of the homology. Even spin structure fall into two classes $\{\Delta_{ia}\}$ and $\{\Delta_{m0}\}$,
according to whether the spin structures on the disconnected components
is even/even or odd/odd,
\bea 
\Delta _{ia} \equiv \left (\matrix{\delta _i \cr \mu _a \cr} \right )
\qquad \qquad 
\Delta _{m0} \equiv \left (\matrix{\nu _m \cr \mu_0 \cr} \right )
\eea
Here, $a=1,2,3$ runs over the even spin structures at genus 1,
$i=0,1,\cdots , 9$ runs over the even spin structures of genus 2 and $m=1,\cdots, 6$ runs over the odd spin structures of genus 2. To save notation, we denote the unique odd spin structure of genus 1 by $\mu_0$, instead of $\nu_0$ as previously. 
The genus 3 $\tet$-constants (non-zero only for even spin structures) have a very simple behavior under the separating degeneration limit:
\bea
\tet [\Delta _{ia} ] (0, \Omega ) 
& \to & \tet _2 [\delta _i] (0, \Omega _2) \cdot \tet _1 [\mu_a]
(0,\tau_3) \equiv \tet _2 [\delta _i] \tet _1 [\mu_a]
\nonumber \\
\tet [\Delta _{m0} ] (0, \Omega ) 
& \to & 0.
\eea
Thus, to study the leading behavior of the $\tet$ constants at genus 3, we shall be interested only in the spin structures $\Delta _{ia}$ and not $\Delta _{m0}$.

\bigskip

\begin{theorem}
Under a separating degeneration into a genus 2 and a genus 1 surface, the limit of the genus 3 form $\Xi_6^\#[\Delta](\Omega^{(3)})$
for an even spin structure $\Delta$ with even/even factorization
on each disconnected component is given by
\be
\label{limit}
\Xi_6^\#[\Delta](\Omega^{(3)})
=
F(\tau)\,
\Xi_6[\delta](\Omega)
+
16\,
\dot
\eta(\tau)^{12}
\,
\bigg\{2\Lambda_A[\delta](\Omega)
+
\Lambda_S[\delta](\Omega)\bigg\}.
\ee
Here $\delta$ is the spin structure component of $\Delta$ on the genus $2$ surface, $F(\tau)$ is 
the genus $1$ form defined by
\be
\label{F}
F(\tau)=\tet_2(\tau)^{12}
+
\tet_3(\tau)^{12}
+
\tet_4(\tau)^{12}
+
3\tet_4(\tau)^4\tet_3(\tau)^8
+
3\tet_3(\tau)^4\tet_2(\tau)^8
+
3\tet_3(\tau)^4\tet_4(\tau)^8
\ee
and $\Lambda_A[\delta](\Omega)$, $\Lambda_S[\delta](\Omega)$ are the genus 2 forms defined both by sums of the form
\be
\sum
\<\delta|\e\>
\<\delta|\eta\>
\<\delta|\kappa\>
\,
\tet[\e]^4(0,\Omega)
\tet[\eta]^4(0,\Omega)
\tet[\kappa]^4(0,\Omega),
\ee
but over different sets of triplets $\{\e,\eta,\kappa\}$ of even spin structures. For $\Lambda_A[\delta](\Omega)$, the triplet
$\{\e,\eta,\kappa\}$ is required to be asyzygous, while for $\Lambda_S[\delta](\Omega)$, it is required to be syzygous. For both, $\{\e,\eta,\kappa\}$ is also required to combine with $\delta$ into a quartet which contains exactly 2 asyzygous triplets. 
\end{theorem} 

\bigskip
\noindent
{\it Proof.}
Under $Sp(6,{\bf Z})$ modular transformations, we can map any even spin structure into $\Delta=(0|0)$,
in which case $\Xi_6[\Delta](\Omega^{(3)})$ becomes 
\bea
\Xi _6^\# [\Delta ] (\Omega^{(3)} ) =
- \half \sum _{[\Delta, \epsilon, \eta, \kappa]}
 \tet [\epsilon] ^4(0,\Omega^{(3)}) \tet [\eta]^4 (0,\Omega^{(3)})
\tet [\kappa]^4 (0,\Omega^{(3)})
\eea
In the degenerating limit, only spin structures of the form $\Delta _{ia}$
will survive, so we may immediately restrict $\epsilon, \eta, \kappa$ to be
of that form,
\bea
\epsilon _{ia} \equiv \left (\matrix{\delta _i \cr \mu _a \cr} \right )
\qquad 
\eta _{jb} \equiv \left (\matrix{\delta _j \cr \mu _b \cr} \right )
\qquad 
\kappa _{kc} \equiv \left (\matrix{\delta _k \cr \mu _c \cr} \right )
\eea
The asymptotics is now manifest,
\bea
-2\Xi _6^\# [\Delta ] (\Omega^{(3)} )  \to 
 \sum _{[\delta_{11}, \epsilon_{ia}, \eta_{jb}, \kappa _{kc}]}
\tet _2 [\delta _i] ^4  \tet _2 [\delta _j ]^4  \tet _2 [\delta _k]^4
\cdot
\tet _1 [\mu _a] ^4  \tet _1 [\mu _b ]^4  \tet _1 [\mu _c]^4
\eea
It remains to identify the totally asyzygous quartets that contribute and to parametrize them
in a convenient way. The conditions are
\bea
e(\delta _1, \delta _i, \delta _j) \cdot e(\mu_1, \mu _a, \mu _b) & = & -1
\nonumber \\
e(\delta _1, \delta _j, \delta _k) \cdot e(\mu_1, \mu _b, \mu _c) & = & -1
\nonumber \\
e(\delta _1, \delta _k, \delta _i) \cdot e(\mu_1, \mu _c, \mu _a) & = & -1
\nonumber \\
e(\delta _i, \delta _j, \delta _k) \cdot e(\mu_a, \mu _b, \mu _c) & = & -1
\eea
It is not hard to analyze the equation on the genus 1 spin structures.
Denoting the choice of the 3 genus 1 spin structures simply by their
numbers $abc$, and listing the 4 values of $e$ above in a row vector, we
have
\bea
\label{foursigns}
111, 222, 333 && ++++ \qquad 123, 132      \quad +-+- \nonumber \\
112, 121, 211 && ++++ \qquad 231, 321      \quad -++- \nonumber \\
113, 131, 311 && ++++ \qquad 312, 213      \quad ++-- \nonumber \\
122, 212, 221 && ++++ \qquad 223, 332      \quad  +--+ \nonumber \\
133, 313, 331 && ++++ \qquad 232, 323      \quad --++ \nonumber \\
              &&  \hskip 1in 322, 233      \quad -+-+ 
\eea
Thus, up to permutations, there are only two cases.
The case where all signs are + produces genus 2 totally asyzygous
quartets of the form $[\delta
_1, \delta _i, \delta _j, \delta _k]$, since we have
\bea
e(\delta _1, \delta _i, \delta _j)  = -1 & \qquad &
e(\mu_1, \mu _a, \mu _b) =+1 
\nonumber \\
e(\delta _1, \delta _j, \delta _k)  = -1 & \qquad &
e(\mu_1, \mu _b, \mu _c) = +1 
\nonumber \\
e(\delta _1, \delta _k, \delta _i)  = -1 & \qquad &
e(\mu_1, \mu _c, \mu _a) = +1 
\nonumber \\
e(\delta _i, \delta _j, \delta _k)  = -1 & \qquad &
e(\mu_a, \mu _b, \mu _c) = +1
\eea
Thus $\{\delta _1, \delta _i, \delta _j, \delta _k\}$ form a 
totally asyzygous quartet, and we recognize the
structure of $\Xi_6[\delta_1](\Omega^{(2)})$.
Furthermore, the fact that they are totally asyzygous implies that these spin structures must all be distinct. The corresponding contributions tend towards 
the first term on the right hand side of
(\ref{limit}).

\medskip

In the remaining contributions, with two $+$ and two $-$ signs
in (\ref{foursigns}), the genus 1 part
is actually always proportional to $\,\eta ^{12}$. This is clear in the first three lines, where the product $\tet _2 ^4  \tet _3 ^4 \tet _4 ^4=
16 \eta ^{12}$ manifestly factors out. But it is also true in the last
three lines, upon adding up the contributions from the two cases listed.
For example,
\bea 
223 \to  \tet _2 ^8 \tet _4 ^4 \hskip 1in
332 \to  \tet _2 ^4 \tet _4 ^8
\eea
Their sum is $\tet _2 ^8 \tet _4 ^4 + \tet _2 ^4 \tet _4 ^8 = \tet _2 ^4
\tet _3 ^4  \tet _4 ^4 = 16 \eta ^{12}$. 
It remains only to identify the genus 2 contribution.
In the present case, the values $e(\delta_1,\delta_i,\delta_j)$,
$e(\delta_1,\delta_j,\delta_k)$, $e(\delta_1,\delta_i,\delta_k)$,
$e(\delta_i,\delta_j,\delta_k)$ must contain exactly two signs $+$ and two signs $-$, i.e.,
the quartet $\{\delta_1,\delta_i,\delta_j,\delta_k\}$
contains exactly two syzygous triplets and two asyzygous triplets.
These quartets come in two groups, one group characterized by
the fact that $\{\delta_i,\delta_j,\delta_k\}$ is an asyzygous triplet, and the other characterized by the fact that
$\{\delta_i,\delta_j,\delta_k\}$ is syzygous. 
Since the first three lines of (\ref{foursigns}) correspond to $\{\delta_i,\delta_j,\delta_k\}$ asyzygous and the next three to
$\{\delta_i,\delta_j,\delta_k\}$ syzygous, Theorem 5 follows.
Q.E.D.

\begin{appendix}

\section{Modular transformations for any genus}
\setcounter{equation}{0}

Modular transformations  in genus $h$ form the infinite discrete
group $Sp(2h,{\bf Z})$, defined by
\be
M=\left ( \matrix{A & B \cr C & D \cr} \right )
\qquad \qquad
\left ( \matrix{A & B \cr C & D \cr} \right ) 
\left ( \matrix{0 & I \cr -I & 0 \cr} \right )
\left ( \matrix{A & B \cr C & D \cr} \right ) ^t
=
\left ( \matrix{0 & I \cr -I & 0 \cr} \right )
\ee
where $A,B,C,D$ are integer valued $h \times h$ matrices and the
superscript ${}^t$ denotes transposition. 
To exhibit the action of the modular group on 1/2 characteristics
$\kappa$ (even or odd), it is convenient to assemble the 1/2
characteristics into a single column of $2h$ entries and the action of the
modular group is then given by \cite{fay}
\be
\left (\matrix{ \tilde \kappa' \cr \tilde \kappa ''\cr}  \right )
=
\left ( \matrix{D & -C \cr -B & A \cr} \right )
\left ( \matrix{ \kappa ' \cr \kappa '' \cr} \right )
+ \half \ {\rm diag} 
\left ( \matrix{CD^T  \cr AB^T \cr} \right )
\ee
Here and below, ${\rm diag} (M)$ of a $h \times h$ matrix $M$ is an
$1\times h$ column vector whose entries are the diagonal entries on $M$. 
On the period matrix, the transformation acts by
\be
\tilde \Omega = (A\Omega + B ) (C\Omega + D)^{-1}
\ee
while on the Jacobi $\tet$-functions, the action is given by
\be
\tet [\tilde \kappa ] \biggl ( \{(C\Omega +D)^{-1} \}^t \zeta , \tilde
\Omega \biggr ) =
\epsilon (\kappa, M) \det (C\Omega + D) ^\half 
e^{ i \pi \zeta ^t (C\Omega +D)^{-1} C \zeta }
\tet [ \kappa ] (\zeta, \Omega)
\ee
where $\kappa = (\kappa ' |\kappa '')$ and $\tilde \kappa = (\tilde
\kappa ' | \tilde \kappa '')$. The phase factor $\epsilon (\kappa, M)$
depends upon both $\kappa $ and the modular transformation $M$ and obeys 
$\epsilon (\kappa , M )^8=1$.

\subsection{Convenient generators of the modular group}

Using a Lemma (15) of \cite{igusa2}, the group is generated by 
\bea
\left ( \matrix{A & B \cr 0 & D \cr} \right )
\qquad {\rm and} \qquad 
\left ( \matrix{0 & I \cr -I & 0 \cr} \right )
\eea
where $I$ is the $h \times h$ identity matrix. The first set is easily
seen to be generated in turn by the subgroups generated by matrices of
the form $M_A$ and $M_B$, so that the full $Sp(2h,{\bf Z})$ is generated
by three subgroups,
\bea
M_A = \left ( \matrix{A & 0 \cr 0 & (A^t)^{-1} \cr} \right )
\qquad 
M_B = \left ( \matrix{I & B \cr 0 & I \cr} \right )
\qquad 
S = \left ( \matrix{0 & I \cr -I & 0 \cr} \right )
\eea
where $A \in SL(h,{\bf Z})$ and $B^t=B$, the latter forming the
translation group ${\bf Z}^{\half h(h+1)}$. The action on spin structures
for each of the 3 subgroups  is given by
\bea
M_A (\kappa '|\kappa '') & = & ((A^t)^{-1} \kappa ' | A
\kappa '')
\nonumber \\
M_B (\kappa '|\kappa '') & = & ( \kappa ' | A \kappa '' - B \kappa ' +
\half {\rm diag} B)
\nonumber \\
S \ (\kappa '|\kappa '') & = & ( \kappa '' |  \kappa ')
\eea

\subsection{Modular transformations of $\tet$-constants}

We shall be most interested in the modular transformations of
$\tet$-constants of even spin structures $\delta$ , which are given by 
\be
\tet [\tilde \delta ] (0  , \tilde \Omega ) =
\epsilon (\delta, M) \det (C\Omega + D) ^\half 
\tet [ \delta ] (0, \Omega)
\ee
and we have 
\bea
\epsilon (\delta, M_A ) & = & 1
\nonumber \\
\epsilon (\delta, M_B ) & = & \exp \{ - i \pi \delta ' {}^t B \delta '
+ i \pi \delta ' {}^t {\rm diag} B \}
\nonumber \\
\epsilon (\delta, S ) & = & \exp \{i \pi h/4 \}
\eea

\subsection{Modular transformations of signatures}

The signature on pairs is {\sl not} invariant under all modular
transformations; instead, 
\bea
\< M_A \delta | M_A \epsilon \> & = & \< \delta | \epsilon \>
\nonumber \\
\< M_B\delta | M_B \epsilon \> & = & \< \delta | \epsilon \> 
\cdot \exp \{2 \pi i (\delta ' - \epsilon ')^t {\rm diag} B\}
\nonumber \\
\< S \delta | S \epsilon \> & = & \< \delta | \epsilon \>
\eea
This transformation law means, however, that the signature for triplets
is invariant under all modular transformations. It suffices to check this
for transformations $M_B$,
\bea
e(M_B \delta, M_B \epsilon, M_B \eta)
& = & \< M_B \delta | M_B \epsilon \> \< M_B \epsilon  | M_B \eta \>    
\< M_B \eta | M_B \delta \>
\nonumber \\
& = & e(\delta, \epsilon, \eta)
\eea
where the exponential factors arising from the transformation law of the
three pairs cancel. 
In particular, totally asyzygous $N$-tuples are mapped into 
totally asyzygous $N$-tuples under modular transformations.

\section{Spin structures and asyzygies in genus 2}

The spin structures for genus 2 Riemann surfaces are described in detail in \cite{IV},
\S 2.1-\S 2.3. For the readers' convenience, we reproduce here
the notation, since it is used extensively below.
The odd spin structures may be labeled by
\bea
\label{listodd}
2\nu_1 =\left (\matrix{0 \cr  1\cr} \bigg | \matrix{0 \cr 1\cr} \right )
\qquad
2\nu_3 =\left (\matrix{0 \cr  1\cr} \bigg | \matrix{1 \cr 1\cr} \right )
\qquad
2\nu_5 =\left (\matrix{1 \cr  1\cr} \bigg | \matrix{0 \cr 1\cr} \right )
\nonumber \\
2\nu_2 =\left (\matrix{1 \cr  0\cr} \bigg | \matrix{1 \cr 0\cr} \right )
\qquad
2\nu_4 =\left (\matrix{1 \cr  0\cr} \bigg | \matrix{1 \cr 1\cr} \right )
\qquad
2\nu_6 =\left (\matrix{1 \cr  1\cr} \bigg | \matrix{1 \cr 0\cr} \right )
\eea
and the even spin structures by
\bea
2\delta_1 =\left (\matrix{0 \cr 0\cr} \bigg | \matrix{0 \cr 0\cr} \right )
\qquad
2\delta_2 =\left (\matrix{0 \cr 0\cr} \bigg | \matrix{0 \cr 1\cr} \right )
\qquad
2\delta_3 =\left (\matrix{0 \cr 0\cr} \bigg | \matrix{1 \cr 0\cr} \right )
\qquad
2\delta_4 =\left (\matrix{0 \cr 0\cr} \bigg | \matrix{1 \cr 1\cr} \right )
\nonumber \\
2\delta_5 =\left (\matrix{0 \cr 1\cr} \bigg | \matrix{0 \cr 0\cr} \right )
\qquad
2\delta_6 =\left (\matrix{0 \cr 1\cr} \bigg | \matrix{1 \cr 0\cr} \right )
\qquad
2\delta_7 =\left (\matrix{1 \cr 0\cr} \bigg | \matrix{0 \cr 0\cr} \right )
\qquad
2\delta_8 =\left (\matrix{1 \cr 0\cr} \bigg | \matrix{0 \cr 1\cr} \right )
\nonumber \\
2\delta_9 =\left (\matrix{1 \cr 1\cr} \bigg | \matrix{0 \cr 0\cr} \right )
\qquad
2\delta_0 =\left (\matrix{1\cr 1\cr} \bigg | \matrix{1\cr 1\cr} \right)
\eea 
Generators of genus 2 modular transformations $M_1, M_2, M_3, 
\Sigma, T, S$ are chosen as they were defined in \cite{IV}.

\subsection{Tables of asyzygies in genus 2}

The properties of asyzygies often require a case by case treatment. 
For genus 2, it is convenient to simply
make full tables of asyzygies, and check the desired properties
by inspection of these tables. In this Appendix, we provide the
tables which have been used in this paper.

\medskip

There are 120 distinct triples of which 60 are syzygous, $e=+1$,
\bea
&&
(123)\quad \ (124)\quad \ (127)\quad \ (128)\quad \ (134)\quad \ 
(135)\quad \ (136)\quad \ (149)\quad \ (140)\quad \ (156)\quad \ 
\no \\ &&
(157)\quad \ (159)\quad \ (168)\quad \ (160)\quad \ (178)\quad \ 
(179)\quad \ (180)\quad \ (190)\quad \ (234)\quad \ (239)\quad \ 
\nonumber \\ &&
(230)\quad \ (245)\quad \ (246)\quad \ (256)\quad \ (258)\quad \ 
(259)\quad \ (267)\quad \ (260)\quad \ (278)\quad \ (270)\quad \
\nonumber \\ &&
(289)\quad \ (290)\quad \ (347)\quad \ (348)\quad \ (356)\quad \ 
(358)\quad \ (350)\quad \ (367)\quad \ (369)\quad \ (378)\quad \ 
\nonumber \\ &&
(379)\quad \ (380)\quad \ (390)\quad \ (456)\quad \ (457)\quad \ 
(450)\quad \ (468)\quad \ (469)\quad \ (478)\quad \ (470)\quad \ 
\nonumber \\ &&
(489)\quad \ (490)\quad \ (579)\quad \ (570)\quad \ (589)\quad \ 
(580)\quad \ (679)\quad \ (670)\quad \ (689)\quad \ (680)
\no
\eea
and 60 are asyzygous, $e=-1$, 
\bea
&&
(125)\quad \ (126)\quad \ (129)\quad \ (120)\quad \ (137)\quad \ 
(138)\quad \ (139)\quad \ (130)\quad \ (145)\quad \ (146)\quad \ 
\nonumber \\ &&
(147)\quad \ (148)\quad \ (158)\quad \ (150)\quad \ (167)\quad \ 
(169)\quad \ (170)\quad \ (189)\quad \ (235)\quad \ (236)\quad \ 
\nonumber \\ &&
(237)\quad \ (238)\quad \ (247)\quad \ (248)\quad \ (249)\quad \ 
(240)\quad \ (257)\quad \ (250)\quad \ (268)\quad \ (269)
\nonumber \\ &&
(279)\quad \ (280)\quad \ (345)\quad \ (346)\quad \ (349)\quad \ 
(340)\quad \ (357)\quad \ (359)\quad \ (368)\quad \ (360)\quad \ 
\nonumber \\ &&
(370)\quad \ (389)\quad \ (458)\quad \ (459)\quad \ (467)\quad \ 
(460)\quad \ (479)\quad \ (480)\quad \ (567)\quad \ (568)\quad \ 
\nonumber \\ &&
(569)\quad \ (560)\quad \ (578)\quad \ (590)\quad \ (678)\quad \ 
(690)\quad \ (789)\quad \ (780)\quad \ (790)\quad \ (890)
\no 
\eea

\subsection{Asyzygous quartets in genus 2}

The list of asyzygous quartets can be easily read off the table
of asyzygous triplets given above. It turns out that there are
15 asyzygous quartets in all. They are given in the table below,
together with their transformations under modular transformations.

\begin{table}[htdp]
\begin{center}
\begin{tabular}{|c||c|c|c|c|c|c|} \hline 
 TAQ & $M_1$  & $M_2$ & $M_3$ & $S$ & $\Sigma$ & $T$
                \\ \hline \hline
              1250 
            & 3460 
            & 2150        
            & 1269 
            & 1520       
            & 1370
            & 1458 
 \\ \hline
              1269 
            & 3459 
            & 1269        
            & 1250
            & 1584       
            & 1389
            & 1467 
 \\ \hline
              1370  
            & 1370 
            & 2480        
            & 1389 
            & 1730       
            & 1250
            & 1398
 \\ \hline
              1389  
            & 1389 
            & 2479       
            & 1370 
            & 1467       
            & 1269
            & 1370 
 \\ \hline
              1458  
            & 2368 
            & 2357        
            & 1467 
            & 1269       
            & 1467
            & 1250 
 \\ \hline
              1467 
            & 2357
            & 2368       
            & 1458 
            & 1389       
            & 1458
            & 1269
 \\ \hline
              2357 
            & 1467 
            & 1458        
            & 2368 
            & 2357       
            & 2357
            & 3459
 \\ \hline
              2368 
            & 1458 
            & 1467        
            & 2357 
            & 5678       
            & 2368
            & 3460
 \\ \hline
              2479 
            & 2479 
            & 1389        
            & 2480 
            & 3459       
            & 3459
            & 2497
 \\ \hline
              2480 
            & 2480 
            & 1370        
            & 2479 
            & 5690       
            & 3460
            & 2480
 \\ \hline
              3459 
            & 1269 
            & 3459        
            & 3460 
            & 2479       
            & 2479
            & 2357
 \\ \hline
              3460 
            & 1250 
            & 3460        
            & 3459 
            & 7890       
            & 2480
            & 2368
 \\ \hline
              5678 
            & 5678 
            & 5678        
            & 5678 
            & 2368       
            & 5678
            & 5690
 \\ \hline
              5690 
            & 5690 
            & 5690        
            & 5690 
            & 2480       
            & 7890
            & 5678
 \\ \hline
             7890  
            & 7890 
            & 7890        
            &  7890
            & 3460       
            & 5690
            & 7890
 \\ \hline
\end{tabular}
\end{center}
\caption{Modular transformations of totally asyzygous quartets (TAQ)}
\label{table:6}
\end{table}

\subsection{Admissible sextets in genus 2}

We consider sextets of even spin structures in genus $2$
satisfying the property listed in Theorem 2: they can be
written as the union of three pairs,
the union of any two pairs forming a totally asyzygous
quartet. For convenience, we sometimes refer to sextets with this
propery as {\it admissible sextets}.
In the table below, we have listed all the 15
sextets in genus $2$ satisfying this property.
The three totally asyzygous quartets that each sextet
contains have been denoted by $A$, $B$, and $C$.
The three pairs can then be recaptured by taking
$A\cap B$, $B\cap C$, and $A\cap C$.

\begin{table}[htdp]
\begin{center}
\begin{tabular}{|c||c|c|c|c|c|c|c|} \hline 
sextet & $M_3$ ~ $\epsilon ^2$  & $S$ & $T$ & $\Sigma$  & 
$A \cap B$ & $B \cap C$ & $C \cap A$
                \\ \hline \hline
         123570
         & 123689 $-$
         & 123570
         & 134589
         & 123570
         &10
         &37
         &25
 \\ \hline
              123689
            & 123570 $-$ 
            & 145678        
            & 134670
            & 123689 
            & 19       
            & 38
            & 26 
 \\ \hline
              124580  
            & 124679 $-$ 
            & 125690        
            & 124580
            & 134670 
            & 48       
            & 20
            & 15 
 \\ \hline
         124679
         &124580 $-$ 
         & 134589
         & 124679
         & 134589
         &16
         &47
         &29
 \\ \hline
              125690 
            & 125690 $+$ 
            & 124580        
            & 145678
            & 137890 
            & 12     
            & 69
            & 50 
 \\ \hline
              134589  
            & 134670 $-$
            & 124679        
            & 123570
            & 124679 
            & 18       
            & 45
            & 39
 \\ \hline
         134670
         & 134589 $-$
         & 137890
         & 123689
         & 124580
         &17
         &46
         &30
 \\ \hline
        137890
        & 137890 $+$
         &134670
         & 137890
         & 125690
         &13
         &89
         &70
 \\ \hline
        145678
         & 145678 $+$
         & 123689
         & 125690
         & 145678
         & 14
         &67
         &58
 \\ \hline
         234579
         & 234680 $-$ 
         & 234579 
         & 234579
         & 234579
         &27
         &49
         &35 
 \\ \hline
              234680 
            & 234579 $-$
            & 567890      
            & 234680
            & 234680
            & 28       
            & 40
            & 36 
 \\ \hline
        235678
         & 235678 $+$
         & 235678
         & 345690
         & 235678
         &23
         &68
         &57
 \\ \hline
         247890
         & 247890 $+$ 
         & 345690 
         & 247890
         & 345690
         &24
         &80
         &79
 \\ \hline
              345690  
            & 345690 $+$ 
            & 247890        
            & 235678 
            & 247890
            & 34      
            & 60
            & 59 
 \\ \hline
        567890
         & 567890 $+$
         & 234680
         & 567890
         & 567890
         &56
         &90
         &78
 \\ \hline
\end{tabular}
\end{center}
\caption{Modular transformations and TAQ  structure of admissible sextets }
\label{table:18}
\end{table}

\medskip

Given a fixed spin structure $\delta$, the admissible sextets
split into two groups $s^c[\delta]$ and $s[\delta]$,
characterized respectively as the sextets which
include and do not include $\delta$.
For each fixed $\delta$, $s[\delta]$ always consists
of 6 sextets, and $s^c[\delta]$ of 9 sextets.
In the table below, the first 6 sextets are the
sextets in $s[\delta_7]$, and the remaining ones are the
sextets in $s^c[\delta_7]$.

\section{Genus 3 spin structures \& modular transf}

Here, we provide a summary of the computer results obtained
on genus 3 spin structures, as well as multiplets of spin structures.
The genus 3 spin structures are labeled as follows,
\bea
\Delta _1 = [0,0,0,0,0,0]       \qquad  
\Delta _{13} = [0,1,0,0,1,0]      \qquad
\Delta _{25} = [1,0,0,1,0,1] 
\no \\
\Delta _2 = [0,0,0,0,0,1]           \qquad  
\Delta _{14} = [0,1,0,1,0,0] \qquad
\Delta _{26} = [1,0,0,1,1,0]
\no \\
\Delta _3 = [0,0,0,0,1,0]           \qquad  
\Delta _{15} = [0,1,0,1,0,1] \qquad
\Delta _{27} = [1,0,1,0,0,0] 
\no \\
\Delta _4 = [0,0,0,1,0,0]           \qquad  
\Delta _{16} = [0,1,0,1,1,0] \qquad
\Delta _{28} = [1,0,1,0,0,1]
\no \\
\Delta _5 = [0,0,0,1,0,1]           \qquad  
\Delta _{17} = [0,1,1,0,0,0] \qquad
\Delta _{29} = [1,0,1,0,1,0]
\no \\
\Delta _6 = [0,0,0,1,1,0]           \qquad  
\Delta _{18} = [0,1,1,0,0,1] \qquad
\Delta _{30} = [1,0,1,1,1,1]
\no \\
\Delta _7 = [0,0,1,0,0,0]           \qquad  
\Delta _{19} = [0,1,1,0,1,0]   \qquad
\Delta _{31} = [1,1,0,0,1,1]
\no \\
\Delta _8 = [0,0,1,0,0,1]           \qquad  
\Delta _{20} = [0,1,1,1,1,1]   \qquad 
\Delta _{32} = [1,1,0,1,1,1]
\no \\
\Delta _9 = [0,0,1,0,1,0]       \qquad  
\Delta _{21} = [1,0,0,0,0,0]   \qquad 
 \Delta _{33} = [1,1,1,0,1,1]
\no \\
\Delta _{10} = [0,0,1,1,1,1]    \qquad 
\Delta _{22} = [1,0,0,0,0,1]   \qquad 
\Delta _{34} = [1,1,1,1,0,0]
\no \\
\Delta _{11} = [0,1,0,0,0,0]    \qquad 
\Delta _{23} = [1,0,0,0,1,0]   \qquad 
\Delta _{35} = [1,1,1,1,0,1]
\no \\
\Delta _{12} = [0,1,0,0,0,1]    \qquad 
\Delta _{24} = [1,0,0,1,0,0]   \qquad
\Delta _{36} = [1,1,1,1,1,0]
\eea
and we take the following basis of genus 3 modular generators,
\bea
B_i = \left ( \matrix{ I & b_i \cr 0 & I \cr} \right )
\hskip .5in
A_i = \left ( \matrix{ a_i & 0 \cr 0 & (a_i ^t)^{-1} \cr} \right )
\hskip .5in
S = \left ( \matrix{ 0 & I \cr -I & 0 \cr} \right )
\eea
where
\bea
b_1 = \left ( \matrix{ 1 & 0 & 0 \cr 0 & 0 & 0 \cr 0 & 0 & 0 \cr} \right )
\hskip .5in 
b_2 = \left ( \matrix{ 0 & 0 & 0 \cr 0 & 1 & 0 \cr 0 & 0 & 0 \cr} \right )
\hskip .5in 
b_3 = \left ( \matrix{ 0 & 0 & 0 \cr 0 & 0 & 0 \cr 0 & 0 & 1 \cr} \right )
\no \\
b_4 = \left ( \matrix{ 0 & 1 & 0 \cr 1 & 0 & 0 \cr 0 & 0 & 0 \cr} \right )
\hskip .5in 
b_5 = \left ( \matrix{ 0 & 0 & 1 \cr 1 & 0 & 0 \cr 0 & 0 & 0 \cr} \right )
\hskip .5in 
b_6 = \left ( \matrix{ 0 & 0 & 0 \cr 0 & 0 & 1 \cr 0 & 1 & 0 \cr} \right )
\no \\
a_1 = \left ( \matrix{ 0 & 1 & 0 \cr -1 & 0 & 0 \cr 0 & 0 & 1 \cr} \right )
\hskip .3in 
a_2 = \left ( \matrix{ 1 & 0 & 0 \cr 0 & 0 & 1 \cr 0 & -1 & 0 \cr} \right )
\hskip .4in 
a_3 = \left ( \matrix{ 0 & 0 & 1 \cr 0 & 1 & 0 \cr -1 & 0 & 0 \cr} \right )
\no \\
a_4 = \left ( \matrix{ 1 & 1 & 0 \cr 0 & 1 & 0 \cr 0 & 0 & 1 \cr} \right )
\hskip .5in 
a_5 = \left ( \matrix{ 1 & 0 & 1 \cr 0 & 1 & 0 \cr 0 & 0 & 1 \cr} \right )
\hskip .5in 
a_6 = \left ( \matrix{ 1 & 0 & 0 \cr 0 & 1 & 1 \cr 0 & 0 & 1 \cr} \right ) 
\eea
The action of all elementary modular generators 
($A_i, B_i$, $i=1,\cdots,6$ and $S$) are given in Table \ref{table:30} below.
The action of the additional generator 
$\Lambda \equiv B_1 S B_1 S B_1$ is also given. 
The generators  $B_4,B_5,B_6,A_1,A_2,A_3,A_4,A_5,A_6,S$, 
and $\Lambda $ form a subgroup leaving $\Delta_1$ invariant.
It is conjectured that these generators span the maximal 
subgroup of $Sp(6,{\bf Z}_2)$ which leaves $\Delta _1$ invariant.

\subsection{Multiplets of spin structures}

Asyzygous combinations play an important role.
An asyzygous triplet $(\lambda _1, \lambda _2, \lambda _3)$
is defined so that $e(\lambda _1, \lambda _2, \lambda _3 )=-1$.
An asyzygous $N$-plet (with $N >3$) is a set of spin structures
such that each triplet of distinct spin structures is asyzygous.
The count is as follows,
\bea
\hbox{asyzygous quartets} & \quad & 5040
\no \\
\hbox{asyzygous quintets} & \quad & 2016
\no \\
\hbox{asyzygous sextets} & \quad & 336
\no \\
\hbox{asyzygous septets} & \quad & 0
\eea

\begin{table}[htdp]
\begin{center}
\begin{tabular}{|c||c|c|c|c|c|c||c|c|c|c|c|c||c||c|} \hline
$\delta$ & 
$B_1$  & $B_2$ & $B_3$ & $B_4$  & $B_5$ & $B_6$ & 
$A_1$  & $A_2$ & $A_3$ & $A_4$  & $A_5$ & $A_6$ & 
$S$ & $\Lambda$ 
                    \\ \hline \hline
1   & 11 & 4 & 2 & 1& 1& 1          & 1& 1& 1 & 1& 1& 1         & 1& 1
                 \\ \hline
2   & 12 & 5 & 1 & 2 & 2 & 2            & 2 & 4 & 11 & 2 & 12 & 2   & 3 & 2
                \\ \hline
3   & 13 & 6 & 3 & 3 & 13 & 6           & 3 & 7 & 21 & 3 & 3 & 9        & 2 & 3              
                \\ \hline
4   & 14 & 1 &  5 & 4 & 4 & 4           & 11 & 2 & 4 & 14 & 4 & 5   & 7 & 4
                \\ \hline
5   & 15 & 2 & 4 & 5 & 5 & 5            & 12 & 5 & 14 & 15 & 15 & 4 & 9 & 5
            \\ \hline
6   & 16 & 3 & 6 & 6 & 16 & 3           & 13 & 8 & 24 & 16 & 6 & 10     & 8 & 6
                \\ \hline
7   & 17 & 7 & 8 & 17 & 7 & 8       & 21 & 3 & 7 & 7 & 7 & 7        & 4 & 7
                \\ \hline
8   & 18 & 8 & 7 & 18 & 8 & 7           & 22 & 6 & 17 & 8 & 18 & 8  & 6 & 8
                \\ \hline
9   & 19 & 9 & 9 & 19 & 19 & 10     & 23 & 9 & 27 & 9 & 9 & 3       & 5 & 9             
            \\ \hline
10  & 20 & 10 & 10 & 20 & 20 & 9        & 31 & 10 & 34 & 20 & 20 & 6    & 10 & 10
        \\ \hline
11  & 1 & 14 & 12 & 11 & 11 & 11        & 4 & 11 & 2 & 11 & 11 & 11 & 21 & 21
        \\ \hline
12  & 2 & 15 & 11 & 12 & 12 & 12        & 5 & 14 & 12 & 12 & 2 & 12     & 23 & 22
        \\ \hline
13  & 3 & 16 & 13 & 13 & 3 & 16     & 6 & 17 & 22 & 13 & 13 & 19    & 22 & 23
        \\ \hline
14  & 4 & 11 & 15 & 14 & 14 & 14        & 14 & 12 & 5 & 4 & 14 & 15     & 27 & 24
        \\ \hline
15  & 5 & 12 & 14 & 15 & 15 & 15        & 15 & 15 & 15 & 5 & 5 & 14     & 29 & 25
        \\ \hline
16  & 6 & 13 & 16 & 16 & 6 & 13     & 16 & 18 & 25 & 6 & 16 & 20    & 28 & 26
        \\ \hline
17  & 7 & 17 & 18 & 7 & 17 & 18     & 24 & 13 & 8 & 17 & 17 & 17    & 24 & 27
        \\ \hline
18  & 8 & 18 & 17 & 8 & 18  & 17        & 25 & 16 & 18 & 18 & 8 & 18    & 26 & 28
        \\ \hline
19  & 9 & 19 & 19 & 9 & 9 & 20      & 26 & 19 & 28 & 19 & 19 & 13 & 25 & 29
        \\ \hline
20  & 10 & 20 & 20 &10 & 10 & 19        & 32 & 20 & 35 & 10 & 10 & 16   & 30 & 30
        \\ \hline
21  & 21 & 24 & 22 & 24 & 22 & 21       & 7 & 21 & 3 & 27 & 23 & 21     & 11 & 11
        \\ \hline
22  & 22 & 25 & 21 & 25 & 21 & 22       & 8 & 24 & 13 & 28 & 31 & 22    & 13 & 12
        \\ \hline
23  & 23 & 26 & 23 & 26 & 31 & 26       & 9 & 27 & 23 & 29 & 21 & 29    & 12 & 13
        \\ \hline
24  & 24 & 21 & 25 & 21 & 25 & 24       & 17 & 22 & 6 & 34 & 26 & 25    & 17 & 14
        \\ \hline
25  & 25 & 22 & 24 & 22 & 24 & 25   & 18 & 25 & 16 & 35 & 32 & 24 & 19 & 15
        \\ \hline
26  & 26 & 23 & 26 & 23 & 32 & 23       & 19 & 28 & 26 & 36 & 24 & 30 & 18 & 16
        \\ \hline
27  & 27 & 27 & 28 & 34 & 28 & 28       & 27 & 23 & 9 & 21 & 29 & 27    & 14 & 17
        \\ \hline
28  & 28 & 28 & 27 & 35 & 27 & 27   & 28 & 26 & 19 & 22 & 33 & 28   & 16 & 18
        \\ \hline
29  & 29 & 29 & 29 & 36 & 33 & 30       & 29 & 29 & 29 & 23 & 27 & 23 & 15 & 19
        \\ \hline
30  & 30 & 30 & 30 & 33 & 36 & 29   & 33 & 30 & 36 & 32 & 35 & 26   & 20 & 20
        \\ \hline
31  & 31 & 32 & 31 & 32 & 23 & 32       & 10 & 34 & 31 & 33 & 22 & 33   & 31 & 31
        \\ \hline
32  & 32 & 31 & 32 & 31 & 26 & 31       & 20 & 35 & 32 & 30 & 25 & 36   & 33 & 32
        \\ \hline
33  & 33 & 33 & 33 &30 & 29 & 36        & 30 & 36 & 33 & 31 & 28 & 31   & 32 & 33
        \\ \hline
34  & 34 & 34 & 35 & 27 & 35 & 35       & 34 & 31 & 10 & 24 & 36 & 35   & 34 & 34
        \\ \hline
35  & 35 & 35 & 34 & 28 & 34 & 34       & 35 & 32 & 20 & 25 & 30 & 34   & 36 & 35
        \\ \hline
36  & 36& 36 & 36 & 29 & 30 & 33        & 36 & 33 & 30 & 26 & 34 & 32 & 35 & 36
 \\ \hline \hline
\end{tabular}
\end{center}
\caption{Modular transformations of genus 3 even spin structures }
\label{table:30}
\end{table}

\end{appendix}


\newpage

\begin{thebibliography}{99}

{\small

\bibitem{I} E.~D'Hoker and D.H.~Phong, ``Two-Loop Superstrings I,
Main Formulas'', Phys. Lett. {\bf B529} (2002) 241-255;
hep-th/0110247.

\bibitem{II} E.~D'Hoker and D.H.~Phong, ``Two-Loop Superstrings II,
The chiral Measure on Moduli Space'', Nucl. Phys. {\bf B636} (2002) 3-60;
hep-th/0110283.

\bibitem{III} E.~D'Hoker and D.H.~Phong, ``Two-Loop Superstrings III,
Slice Independence and Absence of Ambiguities'', Nucl. Phys. {\bf B636}
(2002) 61-79; hep-th/0111016.

\bibitem{IV} E.~D'Hoker and D.H.~Phong, ``Two-Loop Superstrings IV,
The Cosmological Constant and Modular Forms'', Nucl. Phys. {\bf B639}
(2002) 129-181; hep-th/0111040.

\bibitem{dp02}
E.~D'Hoker and D.H.~Phong,
``Lectures on two-loop superstrings",
Hangzhou, Beijing 2002, hep-th/0211111.

\bibitem{igusa1}
J.I. Igusa, ``On the graded ring of
theta constants", Amer. J. Math. {\bf 86} (1964) 219; 
\hfil\break
J.I. Igusa, ``On Siegel modular forms
of genus two", Amer. J. Math. {\bf 84} (1962) 175;
\hfil\break
J.I. Igusa, ``Modular forms and projective
invariants",
Amer. J. Math. {\bf 89} (1967) 817-855.

\bibitem{igusa2}
J.I. Igusa, {\sl Theta Functions}, Springer Verlag, 1972.


\bibitem{Aoki:2003sy}
K.~Aoki, E.~D'Hoker and D.~H.~Phong,
``Two-loop superstrings on orbifold compactifications,''
Nucl.\ Phys.\ B {\bf 688} (2004) 3-69
[arXiv:hep-th/0312181].


\bibitem{dp88} E. D'Hoker and D.H. Phong,
``The geometry of string perturbation theory",
Rev. Modern Physics {\bf 60} (1988) 917-1065.


\bibitem{asyzygyII} E. D'Hoker and D.H. Phong,
``Asyzygies, modular forms, and the superstring measure II",
to appear in hep-th.

\bibitem{KKS}
S. Kachru, J. Kumar, and E. Silverstein,
``Vacuum energy cancellation in a nonsupersymmetric
string", Phys. Rev {\bf D59} (1999) 106004, hep-th/9807076;
\hfil\break
S. Kachru and E. Silverstein,
``Self-dual nonsupersymmetric Type II string compactifications",
JHEP 9811:001 (1998) hep-th/9808056;
\hfil\break
S. Kachru and E. Silverstein,
``On vanishing two loop cosmological constants in
nonsupersymmetric strings",
JHEP 9901:004 (1999) hep-th/9810129.48
\hfil\break
G.~Shiu and S.~H.~H.~Tye,
``Bose-Fermi degeneracy and duality in non-supersymmetric strings,''
Nucl.\ Phys.\ B {\bf 542}, 45 (1999)
[arXiv:hep-th/9808095].

\bibitem{freitag}
E. Freitag,
``{\sl Hilbert modular forms}", Springer-Verlag, Berlin, 1990;
\hfil\break
E. Freitag,
``{\sl Siegelsche Modulfunktionen}",
Springer-Verlag, 1983.

\bibitem{pole}
A. Belavin, A. Polyakov, and A. Zamolodchikov, 
``Infinite conformal symmetry in two-dimensional quantum field theory", Nuclear Phys. B {\bf 241} (1984) 333-380;
\hfil\break
D. Friedan and S. Shenker, 
``The analytic geometry of two-dimensional conformal field theory", Nuclear Phys. B {\bf 281} (1987) 509-545;
\hfil\break
D. Friedan and S. Shenker, 
``The integrable analytic geometry of quantum string", Phys. Lett. B {\bf 175} (1986) 287-296. 



\bibitem{fay} J. Fay {\sl Theta Functions on Riemann surfaces}, Springer
Lecture Notes in Mathematics, No 352, Springer Berlin (1973).




}

\end{thebibliography}
\end{document}